\documentclass[twocolumn]{aastex62}
\usepackage{graphicx}
\usepackage{amssymb}
\usepackage{amsmath}
\usepackage{multirow}
\usepackage{hyperref}

\usepackage{natbib}

\begin{document}


\title{Evidence for a merger induced shock wave in ZwCl\,0008.8+5215 with {\it Chandra} and {\it Suzaku}}

\author{G. Di Gennaro}
\affil{Leiden Observatory, Leiden University, PO Box 9513, 2300 RA Leiden, The Netherlands}
\affil{Harvard-Smithsonian Center for Astrophysics, 60 Garden Street, Cambridge, MA 02138, USA}

\author{R.J. van Weeren}
\affil{Leiden Observatory, Leiden University, PO Box 9513, 2300 RA Leiden, The Netherlands}
\affil{Harvard-Smithsonian Center for Astrophysics, 60 Garden Street, Cambridge, MA 02138, USA}

\author{F. Andrade-Santos}
\affil{Harvard-Smithsonian Center for Astrophysics, 60 Garden Street, Cambridge, MA 02138, USA}

\author{H. Akamatsu}
\affil{SRON Netherlands Institute for Space Research, Sorbonnelaan 2, 3584 CA Utrecht, The Netherlands}

\author{S.W. Randall}
\affil{Harvard-Smithsonian Center for Astrophysics, 60 Garden Street, Cambridge, MA 02138, USA}

\author{W. Forman}
\affil{Harvard-Smithsonian Center for Astrophysics, 60 Garden Street, Cambridge, MA 02138, USA}

\author{R.P. Kraft}
\affil{Harvard-Smithsonian Center for Astrophysics, 60 Garden Street, Cambridge, MA 02138, USA}

\author{G. Brunetti}
\affil{Istituto di Radio Astronomia, INAF, Via Gobetti 101, 40121 Bologna, Italy}

\author{W.A. Dawson}
\affil{Lawrence Livermore National Lab, 7000 East Avenue, Livermore, CA 94550, USA}

\author{N. Golovich}
\affil{Lawrence Livermore National Lab, 7000 East Avenue, Livermore, CA 94550, USA}

\author{C. Jones}
\affil{Harvard-Smithsonian Center for Astrophysics, 60 Garden Street, Cambridge, MA 02138, USA}


\correspondingauthor{Gabriella Di Gennaro}
\email{digennaro@strw.leidenuniv.nl}


\begin{abstract}
We present the results from new deep {\it Chandra} ($\sim410$~ks) and {\it Suzaku} ($\sim180$ ks) observations of the merging galaxy cluster ZwCl\,0008.8+5215 ($z=0.104$). Previous radio observations revealed the presence of a double radio relic located diametrically west and east of the cluster center. 
Using our new {\it Chandra} data, we find evidence for the presence of a shock at the location of the western relic, RW, with a Mach number $\mathcal{M}_{S_X}=1.48^{+0.50}_{-0.32}$ from the density jump. We also measure $\mathcal{M}_{T_X}=2.35^{+0.74}_{-0.55}$ and $\mathcal{M}_{T_X}=2.02^{+0.74}_{-0.47}$  from the temperature jump, with {\it Chandra} and {\it Suzaku} respectively. These values are consistent with the Mach number estimate from a previous study of the radio spectral index, under the assumption of diffusive shock acceleration ($\mathcal{M}_{\rm RW}=2.4^{+0.4}_{-0.2}$). Interestingly, the western radio relic does not entirely trace the X-ray shock. A possible explanation is that the relic traces fossil plasma from nearby radio galaxies which is re-accelerated at the shock. For the eastern relic we do not detect an X-ray surface brightness  discontinuity, despite the fact that radio observations suggest a shock with $\mathcal{M}_{\rm RE}=2.2^{+0.2}_{-0.1}$. The low surface brightness and reduced integration time for this region might have prevented the detection. {\it Chandra} surface brightness profile suggests $\mathcal{M}\lesssim1.5$, while {\it Suzaku} temperature measurements found $\mathcal{M}_{T_X}=1.54^{+0.65}_{-0.47}$.
Finally, we also detect a merger induced cold front on the western side of the cluster, behind the shock that traces the western relic. 
\end{abstract}

\shorttitle{Chandra and Suzaku observations of ZwCl\,0008.8+5215}
\shortauthors{Di Gennaro et al.}
\section{Introduction}
Galaxy clusters grow via mergers of less massive systems in a hierarchical process governed by gravity \citep[e.g.][]{press+schecter74,springel+06}. Evidence of energetic ($\sim 10^{64}$ erg) merger events has been revealed, thanks to the {\it Chandra}'s high-angular resolution (i.e. $0.5^{\prime\prime}$), in the form of sharp X-ray surface brightness edges, namely {\it shocks} and {\it cold fronts} \citep[for a review see][]{markevitch+vikhlinin07}. Both shocks and cold fronts are contact discontinuities, but differ because of the sign of the temperature jump and because the pressure profile is  continuous across a cold front. Moreover, while large-scale shocks are detected only in merging systems \citep[e.g.][]{markevitch+02,markevitch+05,markevitch06,russell+10,macario+11,ogrean+16,vanweeren+17a}, cold fronts have been commonly detected also in cool-core clusters \citep[e.g.][]{mazzotta+01,markevitch+01,markevitch+03,sanders+05,ghizzardi+10}. Shocks are generally located in the cluster outskirts, where the thermal intracluster medium (ICM) emission is faint. Hence they are difficult to detect. Constraints on the shock properties, i.e., the temperature jump,  can be provided by the {\it Suzaku} satellite due to its very low background \citep[though its angular resolution is limited, i.e. 2 arcmin; e.g.][]{akamatsu+15}. 
Other complications arise when shocks and cold fronts are not seen  edge-on, i.e., the merger axis is not perfectly located in the plane of the sky. In such a case, projection effects reduce  the surface brightness jumps, potentially hiding the discontinuity.

Merger events can also be revealed in the radio band, via non-thermal synchrotron emission from diffuse sources not directly related to cluster galaxies. Indeed, part of the energy released by a cluster merger may be used to amplify the magnetic field and to accelerate relativistic particles. Results of such phenomena are the so-called {\it radio relics} and {\it halos}, depending on their position in the cluster and on their morphological, spectral and polarization properties \citep[for reviews see][]{feretti+12,brunetti+jones14}.

ZwCl\,0008.8+5215 \citep[hereafter ZwCl\,0008, $z=0.104$,][]{golovich+17} is an example of galaxy cluster whose merging state was firstly observed in the radio band. Giant Meterwave Radio Telescope (GMRT) at 240 and 610~MHz and the Westerbrook Synthesis Radio Telescope (WSRT) observations in the 1.4~GHz band  revealed the presence of a double radio relic, towards the east and the west of the cluster center \citep{vanweeren+11}.  
The radio analysis, based on the spectral index, suggests a weak shock, with Mach numbers $\mathcal{M}\sim 2$. 
Interestingly, no radio halo has been detected so far in the cluster, despite its disturbed dynamical state \citep{bonafede+17}. A recent optical analysis with the Keck and Subaru telescopes showed a very well defined bimodal galaxy distribution, confirming the hypothesis of a binary merger event \citep{golovich+17}. This analysis, in combination with polarization studies at 3.0 GHz \citep{golovich+17}, 4.85 and 8.35 GHz \citep{kierdorf+17}, and simulations \citep{kang+12}, sets an upper limit to the merger axis of $38^\circ$ with the respect to the plane of the sky.
The masses of the two sub-clusters, obtained via weak lensing analysis, are 
$M_{200,1}=5.73^{+2.75}_{-1.81}\times10^{14}$ M$_\odot$ and $M_{200,2}=1.21^{+1.43}_{-0.63}\times10^{14}$ M$_\odot$, corresponding to a mass ratio of about 5. N-body/hydrodynamical simulations by \cite{molnar+broadhurst17} suggested that the cluster is currently in the outgoing phase, with the first-core crossing occurred less then 0.5~Gyr ago.

The detection of radio relics strongly suggests the presence of shock fronts \citep[e.g.][]{giacintucci+08,vanweeren+10,vanweeren+11b,degasperin+15,pearce+17}. A previous shallow (42~ks) {\it Chandra} observation revealed the disturbed morphology of the ICM, but could not unambiguously confirm the presence of shocks \citep{golovich+17}. In this paper, we present results from deep {\it Chandra} observations, totaling $\sim410$~ks, of the galaxy cluster. We also complement the analysis with {\it Suzaku} observations, totaling $\sim183$ ks.

The paper is organized as follows: in Sect. \ref{sec:obs} we describe the {\it Chandra} and {\it Suzaku} observations and data reduction; a description of the X-ray morphology and temperature map of the cluster, based on the {\it Chandra} observations, are provided in Sect. \ref{sec:res}; X-ray surface brightness profiles and temperature measurements are presented in Sect. \ref{sec:R}.
We end with a discussion and a summary in Sects. \ref{sec:disc} and \ref{sec:conc}. Throughout the paper, we assume a standard $\Lambda$CDM cosmology, with $H_0 = 70$ km s$^{-1}$ Mpc$^{-1}$, $\Omega_m = 0.3$ and $\Omega_\Lambda = 0.7$. This translates to a luminosity distance of $D_{\rm L}=483.3$ Mpc, and a scale of 1.85 $\rm{kpc}/^{\prime\prime}$ at the cluster redshift, $z = 0.104$. All errors are given as $1\sigma$. 

\begin{figure*}
\centering
{\includegraphics[width=0.49\textwidth]{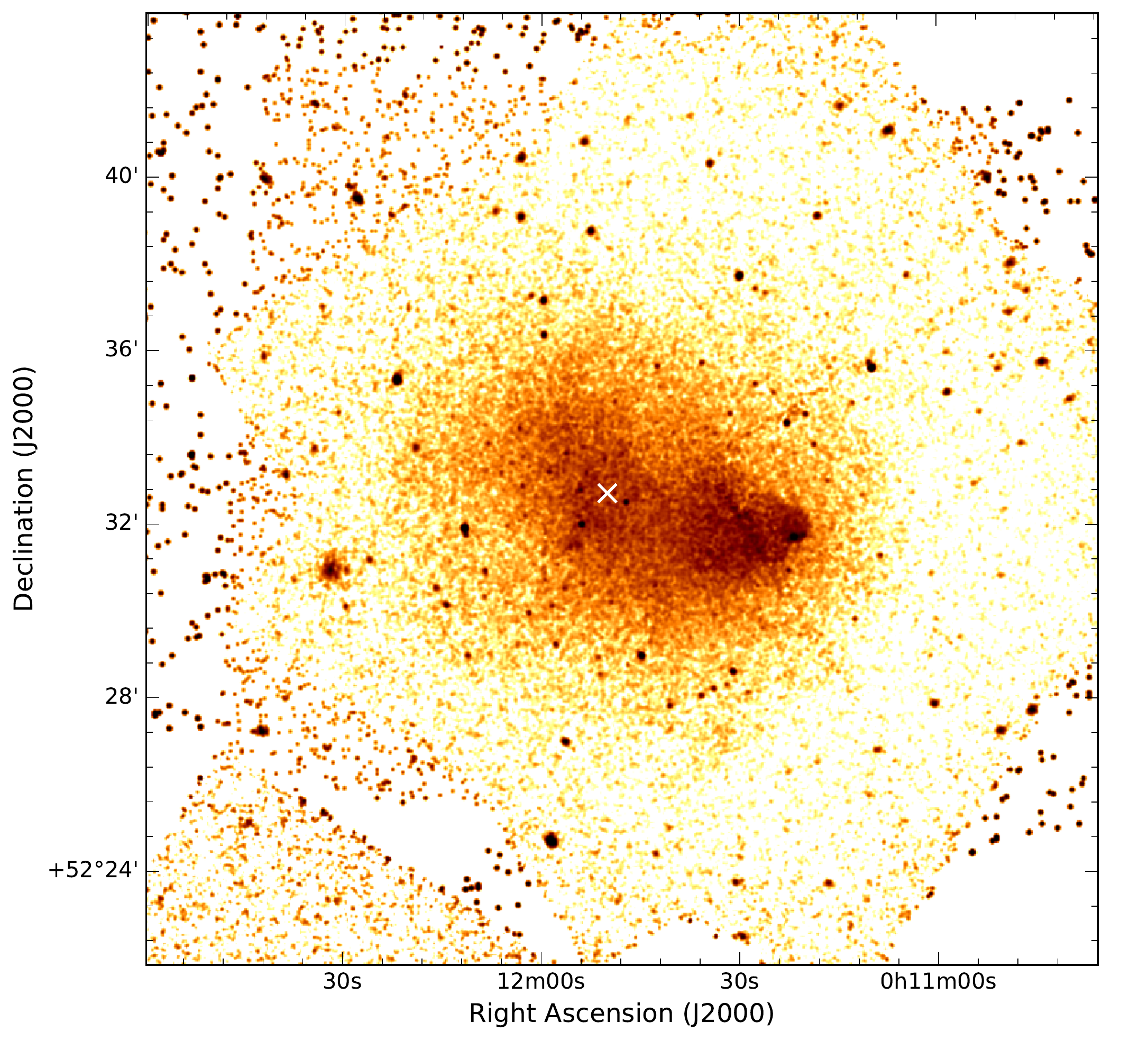}}
{\includegraphics[width=0.49\textwidth]{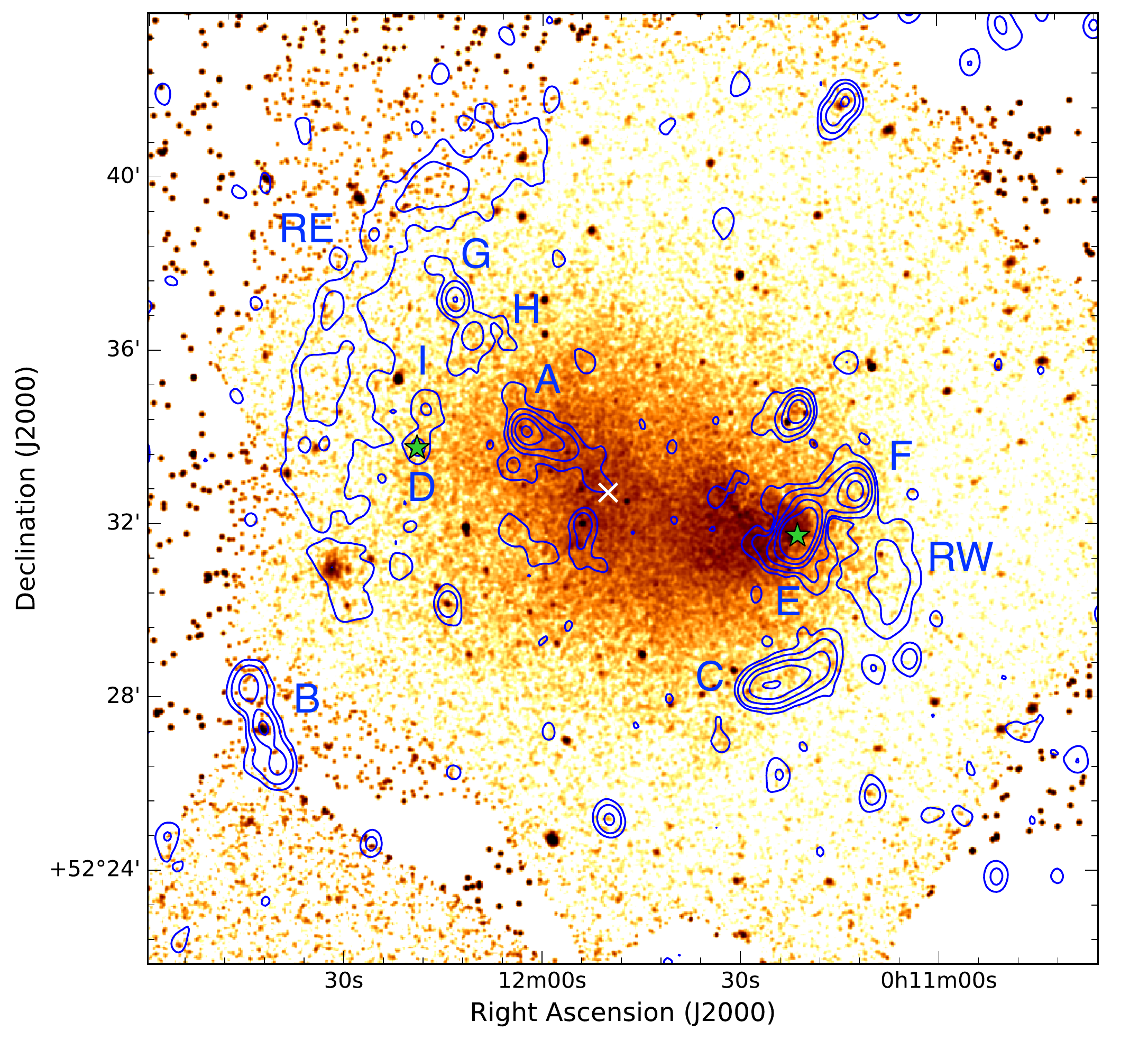}}
\caption{Left panel: background-subtracted, vignetting- and exposure-corrected 0.5--2.0 keV {\it Chandra} image of ZwCl\,0008 smoothed with a 2D Gaussian with $\sigma=2^{\prime\prime}$ (i.e. 1 image pixel). Right panel: the same as the left panel with the 1.4 GHz WSRT radio contours at $4\sigma_{\rm rms}\times[1, 4, 16,...]$ overlaid; the noise level of the radio map is $\sigma_{\rm rms}=27~\mu$Jy beam$^{-1}$ \citep{vanweeren+11}. Radio sources in the right panel have been labeled following \cite{vanweeren+11}, and the two bright central galaxies (BCGs) are identified by the two green stars. The cluster center is identified in the two panels by the white cross.}\label{fig:chandra}
\end{figure*}

\section{Observations and data reduction}\label{sec:obs}
\subsection{{\it Chandra} observations}
We observed ZwCl\,0008 with the Advanced CCD Imaging Spectrometer (ACIS) on {\it Chandra} between 2013 and 2016 for a total time of 413.7 ks. The observation was split into ten single exposures (see the ObsIDs list in Table~\ref{tab:obsid}). The data were reduced using the \texttt{chav} software package\footnote{\url{http://hea-www.harvard.edu/\textasciitilde alexey/CHAV/}} with \texttt{CIAO v4.6} \citep{fruscione+06}, following the processing described in \cite{vikhlinin+05} and applying the \texttt{CALDB v4.7.6} calibration files. This processing includes the application of gain maps to calibrate photon energies, filtering out counts with ASCA grade 1, 5, or 7 and bad pixels, and a correction for the position-dependent charge transfer inefficiency (CTI). Periods with count rates with a factor of 1.2 above and 0.8 below the mean count rate in the 6--12 keV band were also removed. 
Standard blank-sky files were used for background subtraction. The resulting filtered exposure time is 410.1 ks (i.e. 3.6 ks were discarded). 

The final exposure corrected image was made in the 0.5--2.0~keV band by combining all the ObsIDs and using a pixel binning of a factor of four, i.e. $2^{\prime\prime}$. Compact sources were detected in the 0.5--7.0~keV band with the \texttt{CIAO} task \texttt{wavdetect} using scales of 1, 2, 4, 8 pixels and cutting at the $3\sigma$ level. Those compact sources were removed from our spectral and spatial analysis.

\begin{table}
\caption{{\it Chandra} ObsIDs list.}
\begin{center}
\resizebox{0.78\textwidth}{!}{\begin{minipage}{\textwidth}
\begin{tabular}{ccccc}
\hline
\hline
ObsID & Obs. Date & CCD on &Exp. Time & Filtered Exp. Time \\
	& [yyyy-mm-dd] & & [ks] & [ks] \\
\hline
15318 & 2013-06-10 & 0, 1, 2, 3, 6 & 29.0 & 28.9\\
17204 &2015-03-27 & 0, 1, 2, 3, 6, 7 &6.4 & 5.6\\
17205 & 2015-03-17 & 0,1, 2, 3, 6, 7 &6.4 & 5.9\\
18242 & 2016-11-04 & 0, 1, 2, 3 & 84.3 & 83.9\\
18243 & 2016-10-26 & 0, 1, 2, 3, 6, 7 & 30.6 & 30.2 \\
18244 & 2016-10-22 & 0, 1, 2, 3, 6 & 31.7 & 31.2 \\
19901 & 2016-10-17 & 0, 1, 2, 3, 6 & 31.8 & 31.5\\
19902 & 2016-10-19 & 0, 1, 2, 3, 6 & 65.7 & 65.2\\
19905 & 2016-10-29 & 0, 1, 2, 3, 6, 7 & 37.8 & 37.6 \\
19916 & 2016-11-05 & 0, 1, 2, 3 & 90.2 & 90.1 \\
\hline
\end{tabular}
\end{minipage}}
\end{center}
{Note: CCD from 0 to 3: ACIS-I; CCD from 4 to 9: ACIS-S. Back Illuminated (BI) chips: ACIS-S1 and ACIS-S3 (CCD 5 and 7, respectively).}		
\label{tab:obsid}
\end{table}

\begin{table}
\caption{{\it Suzaku} observations and exposure times.}
\begin{center}
\resizebox{0.9\textwidth}{!}{\begin{minipage}{\textwidth}
\begin{tabular}{cccc}
\hline
\hline
Sequence ID & Obs. Date &  Exp. Time & Filtered Exp. Time \\
   & [yyyy-mm-dd] & [ks] & [ks] \\
\hline
809118010 & 2014-07-06 & 119.8 & 98.6\\
809117010 & 2014-07-09 & 102.3 & 84.6 \\
\hline
\end{tabular}
\end{minipage}}
\end{center}		
\label{tab:obsid_suzaku}
\end{table}

\subsection{{\it Suzaku} observations}
{\it Suzaku} observations of ZwCl\,0008 were taken on 6 and 9 July 2014, with two different pointings, to the east  to the west of the cluster center (IDs: 809118010 and 809117010, respectively; see Table~\ref{tab:obsid_suzaku}). Standard data reduction has been performed: data-screening and cosmic-ray cut-off ri
gidity (COR2 $>$ 6 GV to suppress the detector background have been applied \citep[see][for a detailed description of the strategy]{akamatsu+15,akamatsu+17,urdampilleta+18}. We made use of the high-resolution {\it Chandra} observation for the point source identification.
The final cleaned exposure times are 99 and 85~ks (on the east and west pointing, respectively).

\section{Results}\label{sec:res}
\subsection{Global properties}\label{sec:global}
In the left panel in Fig. \ref{fig:chandra}, we present the background-subtracted, vignetting- and exposure-corrected 0.5--2.0 keV {\it Chandra} image of ZwCl\,0008.

\begin{figure*}
\centering
{\includegraphics[width=0.49\textwidth]{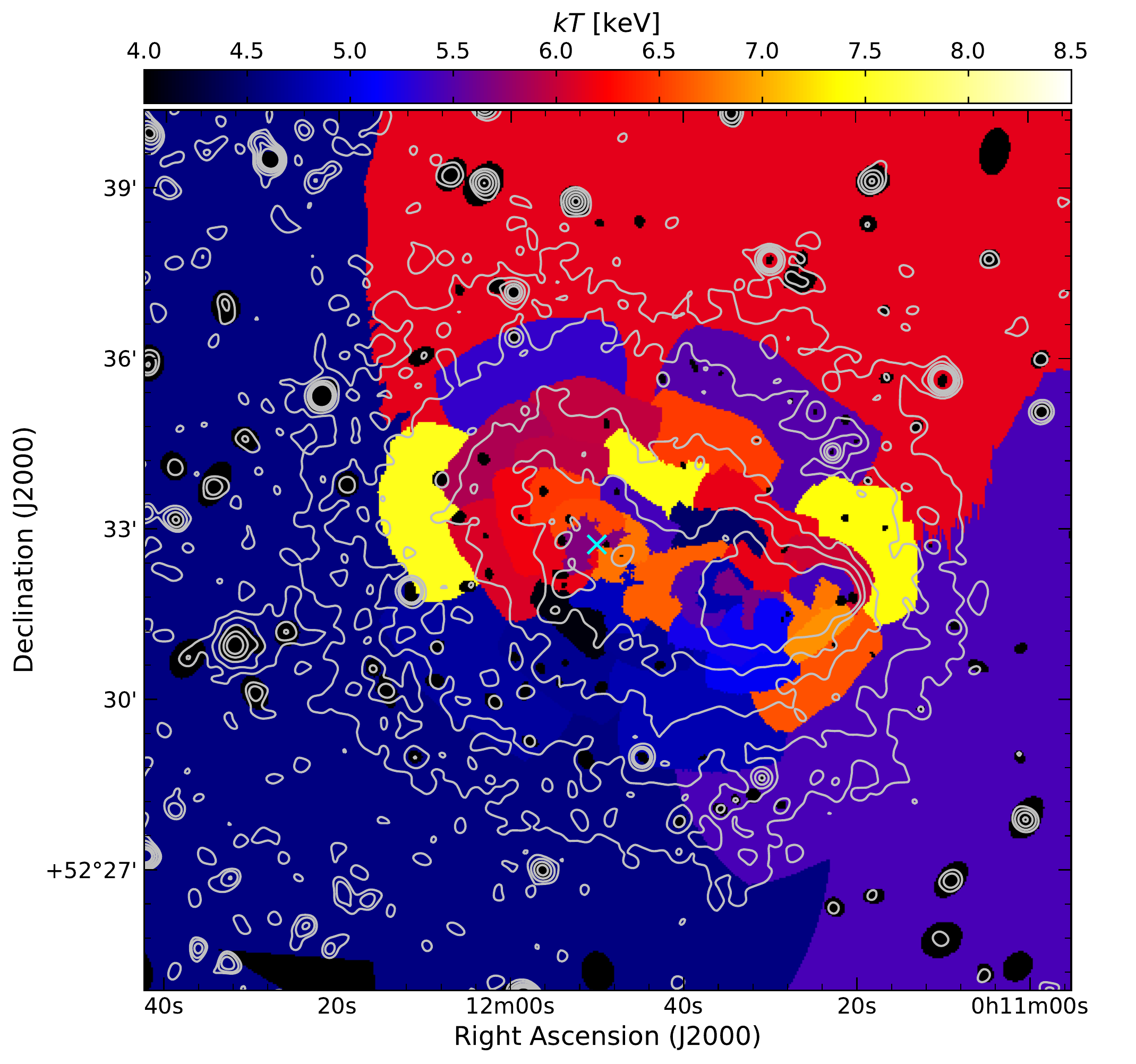}}
{\includegraphics[width=0.49\textwidth]{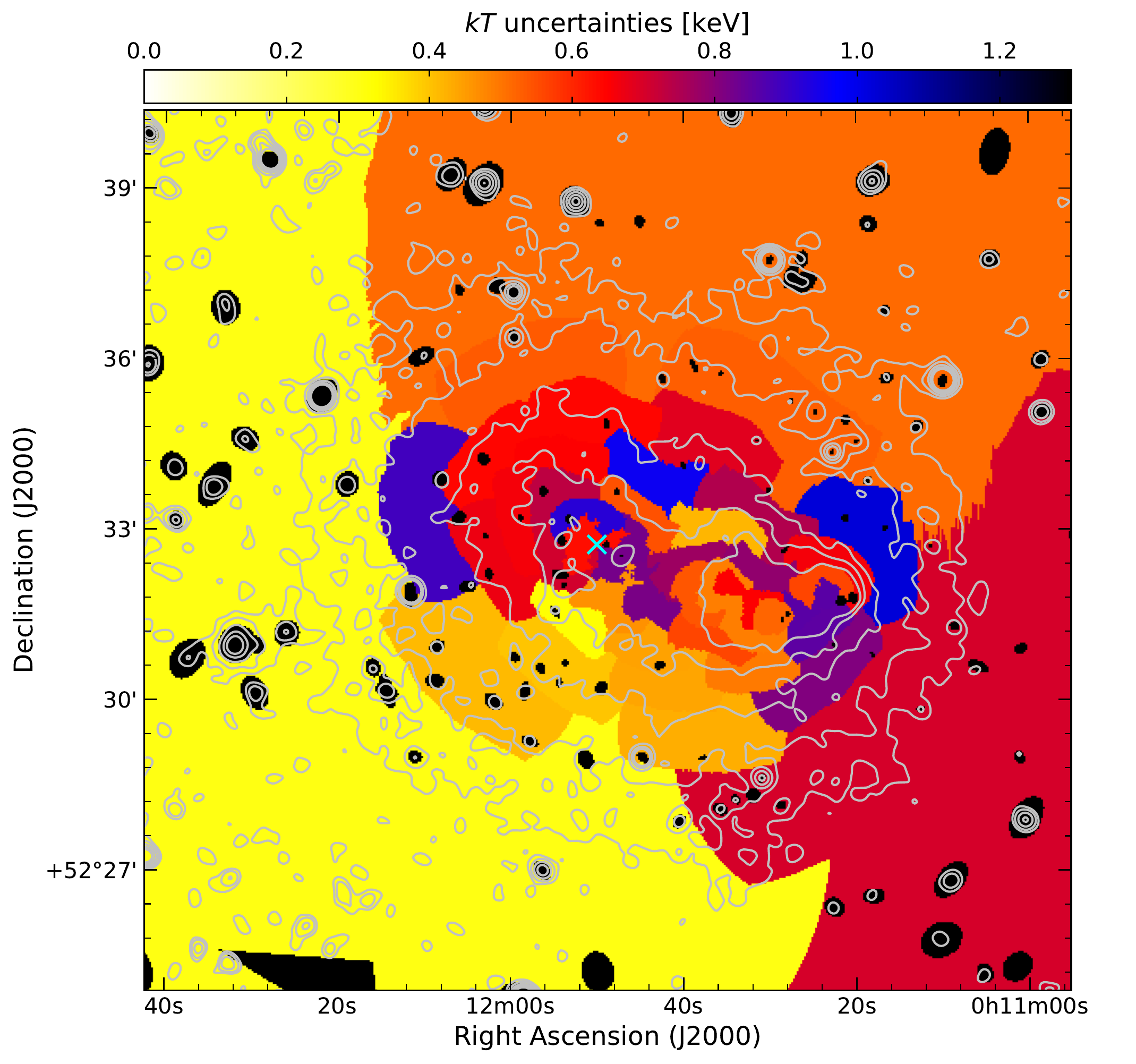}}
\caption{Temperature map (left) and the relative uncertainties (right) of ZwCl\,0008. Each region has a $\rm S/N=40$. Black ellipses represent the compact sources excluded from our spectral and spatial analysis. The cyan cross displays the cluster center (as Fig. \ref{fig:chandra}). X-ray contours (gray) are drawn at $[1.2, 2.4, 4.8, 8.2, 12]\times 10^{-6}$ photons cm$^{-2}$ s$^{-1}$.}\label{fig:T_map}
\end{figure*}

The X-ray emission shows a particularly disturbed morphology: it is elongated from east to west, confirming the merger scenario proposed in the previous studies \citep[e.g.][]{vanweeren+11,golovich+17,molnar+broadhurst17}. The bright, dense remnant core originally associated with the wester BCG lies westward from the cluster center\footnote{The cluster center is taken to be equidistant between the two BCGs, i.e. ${\rm RA=0^h11^m50^s.024}$ and ${\rm DEC=+52^\circ32^\prime37^{\prime\prime}.98}$, J2000 (see white cross in Fig.~\ref{fig:chandra}).}. It has been partly stripped of its material forming a tail of gas towards the north-east. 
It appears to have substantially disrupted the ICM of the eastern sub-cluster and shows a sharp, bullet-like surface brightness edge, 
similarly to the one found in the Bullet Cluster \citep{markevitch+02,markevitch06} and in Abell\,2126 \citep{russell+10,russell+12}. As was also pointed out by \cite{golovich+17}, the remnant core is also coincident with the BCG of the western sub-cluster (marked by a green star symbol in the right panel of Fig.~\ref{fig:chandra}). This is not the case for the eastern sub-cluster's BCG, which is clearly  offset from the X-ray peak (green star in the east in the right panel in Fig.~\ref{fig:chandra}). A  surface brightness discontinuity, extending  about 1~Mpc,  is seen in the western part of the cluster (left panel in Fig.~\ref{fig:chandra}). The location of the western edge is coincident with one of the two radio relics previously detected. However, this relic \citep[hereafter RW,][]{vanweeren+11}  
appears to have a much smaller extent than the X-ray discontinuity. 
To the east, the other radio relic \citep[hereafter RE,][]{vanweeren+11} is located, symmetrically to RW with respect to the cluster center.
This relic is $\sim1.4$ Mpc long, but no clear association with an X-ray discontinuity has been found (see right panel in Fig.~\ref{fig:chandra}).

\begin{figure*}
\centering
\includegraphics[width=\textwidth]{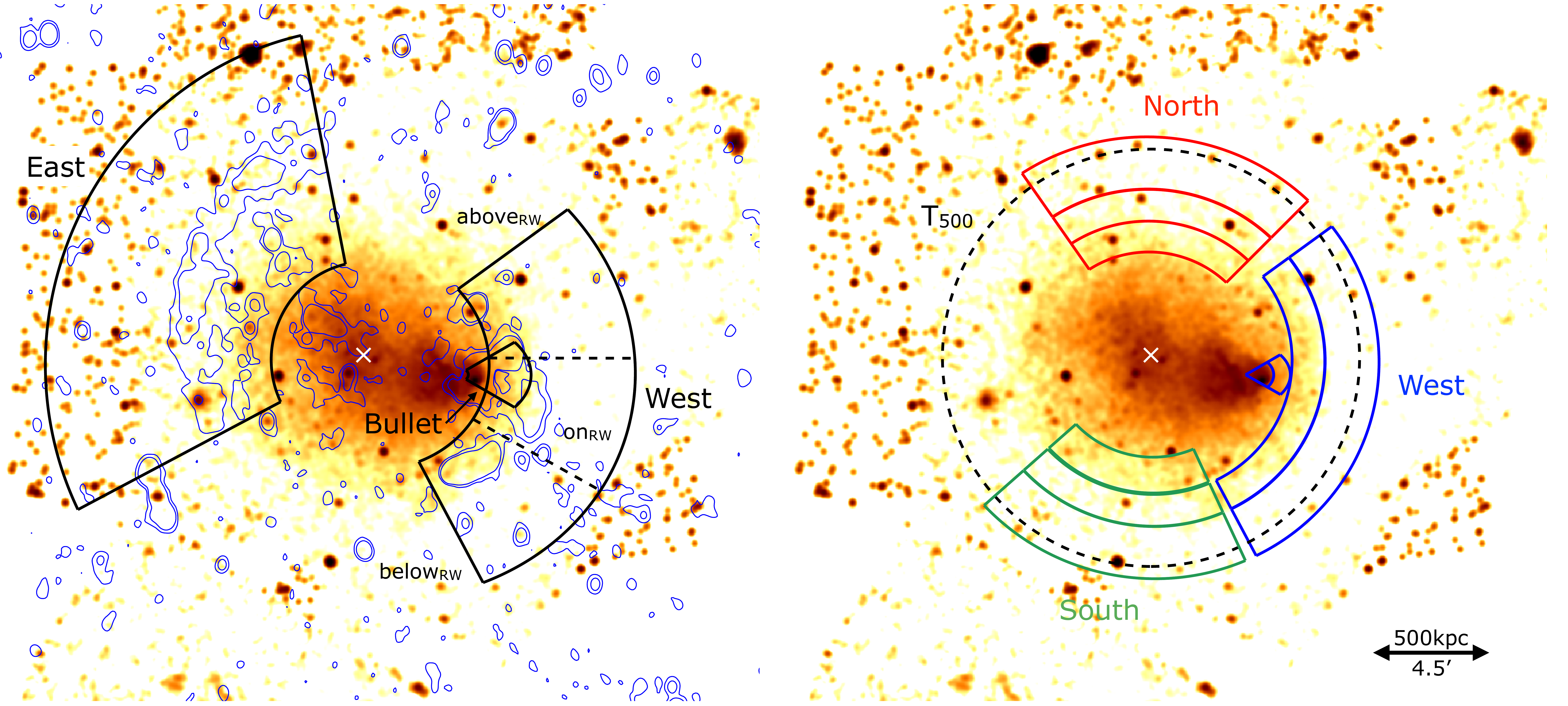}
\caption{Smoothed ($4^{\prime\prime}$) {\it Chandra} 0.5--2.0~keV images showing the sectors used for extracting the surface brightness (left panel) and temperature (right panel) profiles shown in Figs.~\ref{fig:sb_west}, \ref{fig:sb_east}, \ref{fig:sb_cf} and~\ref{fig:temp}. The dashed lines in the left panel show the division for the western edge (sub-sectors above, on, and below the western relic). 
Radio contours in the same panel are drawn at $[1,4]\times3\sigma_{\rm rms}$ levels ($\sigma_{\rm rms}$ is the same as that used in the bottom right panel in Fig.~\ref{fig:chandra}). The black dashed circle in the right panel represents the $R_{500}$ region, from which the cluster average temperature has been obtained. The white cross represents the cluster center.}\label{fig:SB-T_sectors}
\end{figure*}

We determined the X-ray properties of the whole cluster by extracting the spectrum from a circular region with a radius of 0.9~Mpc \citep[approximately $R_{500}$, see][]{golovich+17} centered between the two BCGs (see the black dashed circle in the right panel in Fig.~\ref{fig:SB-T_sectors}).
The cluster spectrum was fitted in the 0.7--7.0~keV energy band with \texttt{XSPEC~v12.9.1u} \citep{arnaud96}. We used a \texttt{phabs*APEC} model, i.e. a single temperature \citep{smith+01} plus the absorption from the hydrogen column density ($N_{\rm H}$) of our Galaxy. We fixed the abundance to $A=0.3$ Z$_\odot$ \citep[abundance table of][]{lodders+09} and $N_{\rm H}=0.311\times10^{22}$ cm$^{-2}$\footnote{Calculation from \url{http://www.swift.ac.uk/analysis/nhtot/}}. The value of Galactic absorption takes the total, i.e. atomic (HI) and molecular (H$_2$), hydrogen column density into account \citep{willingale+13}.
Due to the large number of counts in the cluster, the spectrum was grouped to have a minimum of 50~counts per bin, and the $\chi^2$ statistic was adopted. Standard blank-sky background was used and subtracted from the spectrum of each ObsID.

We found a global cluster temperature and an unabsorbed luminosity\footnote{Since we are fitting simultaneously different ObsID observations, we use the longest exposure ObsID (i.e. 19916, see Table~\ref{tab:obsid}) to obtain the cluster luminosity.} of $kT_{500}=4.83\pm0.06$~keV and 
$L_{\rm [0.1-2.4~keV],500}=1.12\pm0.09\times10^{44}$ erg~s$^{-1}$, respectively. We also repeated the fit, leaving $N_{\rm H}$ free to vary (while the abundance was kept fixed). A resulting temperature of $kT_{500}=4.50\pm0.10$~keV  and column density of $N_{\rm H}=0.342\pm0.007\times10^{22}$ cm$^{-2}$ were found, consistent with the previous results. Our analysis also agree with the results by \cite{golovich+17}\footnote{\cite{golovich+17} found $kT_{500}=4.9\pm0.13$~keV, using $A=0.3$ Z$_\odot$ and $N_{\rm H}=0.201\times10^{22}$ cm$^{-2}$ fixed, with $N_{\rm H}$ the weighted average value from the Leiden/Argentine/Bonn (LAB) survey \citep{kalberla+05}.}.

\subsection{Temperature map}\label{sec:T_map}
We used \texttt{CONTBIN} \citep{sanders06} to create the temperature map of ZwCl\,0008. 
We divided the cluster into individual regions with a signal-to-noise ratio (S/N) of 40. As for the calculation of the global temperature, we removed the contribution of the compact sources, and performed the fit with \texttt{XSPEC12.9.1u} in the 0.7--7.0 keV energy band. The same parameters as in Sect. \ref{sec:global} were used (i.e. $A=0.3$ Z$_\odot$ and $N_{\rm H}=0.311\times10^{22}$ cm$^{-2}$), and we assumed $\chi^2$ statistics. The resulting temperature map, and the corresponding uncertainties, are displayed in Fig.~\ref{fig:T_map} (left and right panel, respectively). 

The disturbed morphology of the cluster is highlighted by the temperature variation in the different regions. Overall, we found that the southeastern part of the cluster appears to have lower temperatures than the northwestern one ($kT_{\rm SE}\sim4.5$ keV and $kT_{\rm NW}\sim6.5$ keV). We measure a region of cold gas ($kT\sim5.5$ keV), in coincidence with the bullet, and a hot region ($kT\sim7.0$ keV) ahead of it, westward in the cluster outskirts. This signature is suggestive of the presence of a cold front. Unfortunately, the S/N required for the temperature map is too high for the identification of any discontinuity at the location of the western outermost edge we see in Fig. \ref{fig:chandra}. Additional hot regions ($kT\sim7.5$ keV) are found  eastward and northwestward of the cluster center.

\section{A search for shocks and cold front}\label{sec:R}
\subsection{Characterization of the discontinuities}\label{sec:discontinuities}
The X-ray signatures described in Section~\ref{sec:global}, and displayed in Fig.~\ref{fig:chandra}, are characteristic of a cluster merger event. 
To confirm the presence of surface brightness discontinuities, we analyzed the surface brightness profile in sectors around the relics. We assume that the X-ray emissivity is only proportional to the density squared ($S_X \propto n^2$), and that the underlying density profile is modeled by a broken power-law model \citep[and references therein]{markevitch+vikhlinin07}:

\begin{align}
n(r) =
\begin{cases}
\mathcal{C}n_0 \bigg (\dfrac{r}{r_{\rm edge}} \bigg )^{-\alpha_1} \, ,\quad &r \leq r_{\rm edge}
\\[12pt]
n_0 \bigg (\dfrac{r}{r_{\rm edge}} \bigg )^{-\alpha_2} \, ,\quad &r > r_{\rm edge} \, .
\end{cases}
\end{align}

Here, $\mathcal{C}\equiv n_1/n_2$ is the compression factor at the jump position (i.e. $r_{\rm edge}$), $n_0$ the density immediately ahead of the putative outward-moving shock front, and $\alpha_1$ and $\alpha_2$ the slopes of the power-law fits. Throughout this paper, the subscripts 1 and 2 are referred to the region behind and ahead the discontinuity (see the right panel in Fig. \ref{fig:SB-T_sectors}), namely the down- and up-stream regions, respectively.
All parameters are left free to vary in the fit. The model is then integrated along the line of sight, assuming spherical geometry and with the instrumental and sky background subtracted. The areas covered by compact sources were excluded from the fitting (see Sect. \ref{sec:obs}). The strongest requirement for the surface brightness analysis is the alignment of the sectors to match the curvature of the surface brightness discontinuities. For this purpose, elliptical sectors\footnote{The ``ellipticity'' of the sector, $e$, is defined as the ratio of the maximum and minimum radius (see Table~\ref{tab:pyxel_res}).} with different aperture angles have been chosen (see the left panel in Fig.~\ref{fig:SB-T_sectors}). The adopted minimum number of required counts per bin are listed in Table~\ref{tab:pyxel_res}.
 
According to this model, a surface brightness discontinuity is detected when $\mathcal{C}>1$, meaning that in the downstream region, i.e. $r \leq r_{\rm edge}$, the gas has been compressed. In the case of a shock, there is a relation between the compression factor $\mathcal{C}$ and the Mach number ($\mathcal{M}=v_{\rm shock}/c_s$, where $v_{\rm shock}$ is the velocity of the pre-shock gas and $c_s$ the sound velocity in the medium\footnote{$c_s=\sqrt{\dfrac{\gamma k T_2}{\mu m_{\rm H}}}$, where $k$ is the Boltzmann constant, $\gamma$ the adiabatic index, $\mu=0.6$ the mean molecular weight and $m_{\rm H}$ the proton mass. $kT_2$ is the pre-shock, i.e. unperturbed medium, temperature.}), via the Rankine-Hugoniot relation \citep{landau+lifshitz59}:

\begin{equation}\label{eq:mach_c}
\mathcal{M}_{S_X}=\sqrt{\frac{2\mathcal{C}}{\gamma+1-\mathcal{C}(\gamma-1)}} + syst_{S_X}\, ,
\end{equation}
where $\gamma$ is the adiabatic index of the gas, and is assumed to be $5/3$ (i.e. a monoatomic gas). The parameter $syst_{S_X}$ takes all the unknown uncertainties into account, e.g. projection effects, curvature of the sector, background estimation, etc. Unfortunately, all these parameters are not easily quantified, so they are embedded in the assumption of our model.

The surface brightness analysis has been performed with \texttt{PyXel}\footnote{\url{https://github.com/gogrean/PyXel}} \citep{ogrean17}, and the uncertainties on the best-fitting parameters are determined using a Markov chain Monte Carlo (MCMC) method \citep{foreman-mackey+13}.

The nature of the confirmed X-ray surface discontinuities is determined by the analysis of the temperature ratio of the down- and up-stream regions, in correspondence of the edge. Shocks and cold fronts are defined to have $T_1/T_2>1$ and $T_1/T_2<1$, respectively \citep{markevitch+vikhlinin07}. For a cold front, the jump in temperature has similar, but inverse, amplitude to the density compression. Hence, they are also characterized by pressure equilibrium across the discontinuity (i.e. $P_1/P_2=1$\footnote{$P=kn_{\rm e}T$, with $k$ the Boltzmann constant and $n_{\rm e}$ the electron density.}). In case of shock front, the Rankine-Hugoniot jump conditions relate the temperature jump, $\mathcal{R}\equiv T_1/T_2$, to the Mach number 
\citep[e.g.][]{landau+lifshitz59}:
\begin{equation}\label{eq:mach_t}
\mathcal{M}_{T_X}= \sqrt{\frac{(8\mathcal{R}-7) + \sqrt{(8\mathcal{R}-7)^2+15} }{5}} + syst_{T_X}\quad ,
\end{equation}
where $\gamma=5/3$ has been used, as for Eq.~\ref{eq:mach_c}. Again, $syst_{T_X}$ takes all the unknown temperature-related uncertainties into account, such as the variation of the metal abundance ($A$) and the Galactic absorption ($N_{\rm H}$) towards the cluster outskirts, background subtraction, etc. \citep[for a more extensive description of the possible systematic uncertainties see][]{akamatsu+17}.

Sectors for the radial temperature measurements have been chosen similarly to the ones used for the surface brightness analysis (see the right panel in Fig.~\ref{fig:SB-T_sectors}), which also provides the accurate position of the edges. 
As for the global cluster analysis (see Sect.~\ref{sec:global}), we fit each spectrum with a single temperature, taking into account the Galactic absorption (\texttt{phabs*apec}). Both the abundance and hydrogen column density were fixed, at $A=0.3$ Z$_\odot$ and $N_{\rm H}=0.311\times10^{22}$ cm$^{-2}$, respectively. Since the number of counts in cluster outskirts are usually low, the spectrum was grouped to have a minimum of 1 count per bin, and the Cash statistic \citep{cash79} was adopted. The ACIS readout artifacts
were not subtracted in our analysis. This does not affect the analysis, since the cluster is relatively faint and no bright compact source is contaminating the observations.

\begin{figure}
\centering
\vspace{5mm}
\includegraphics[width=0.45\textwidth]{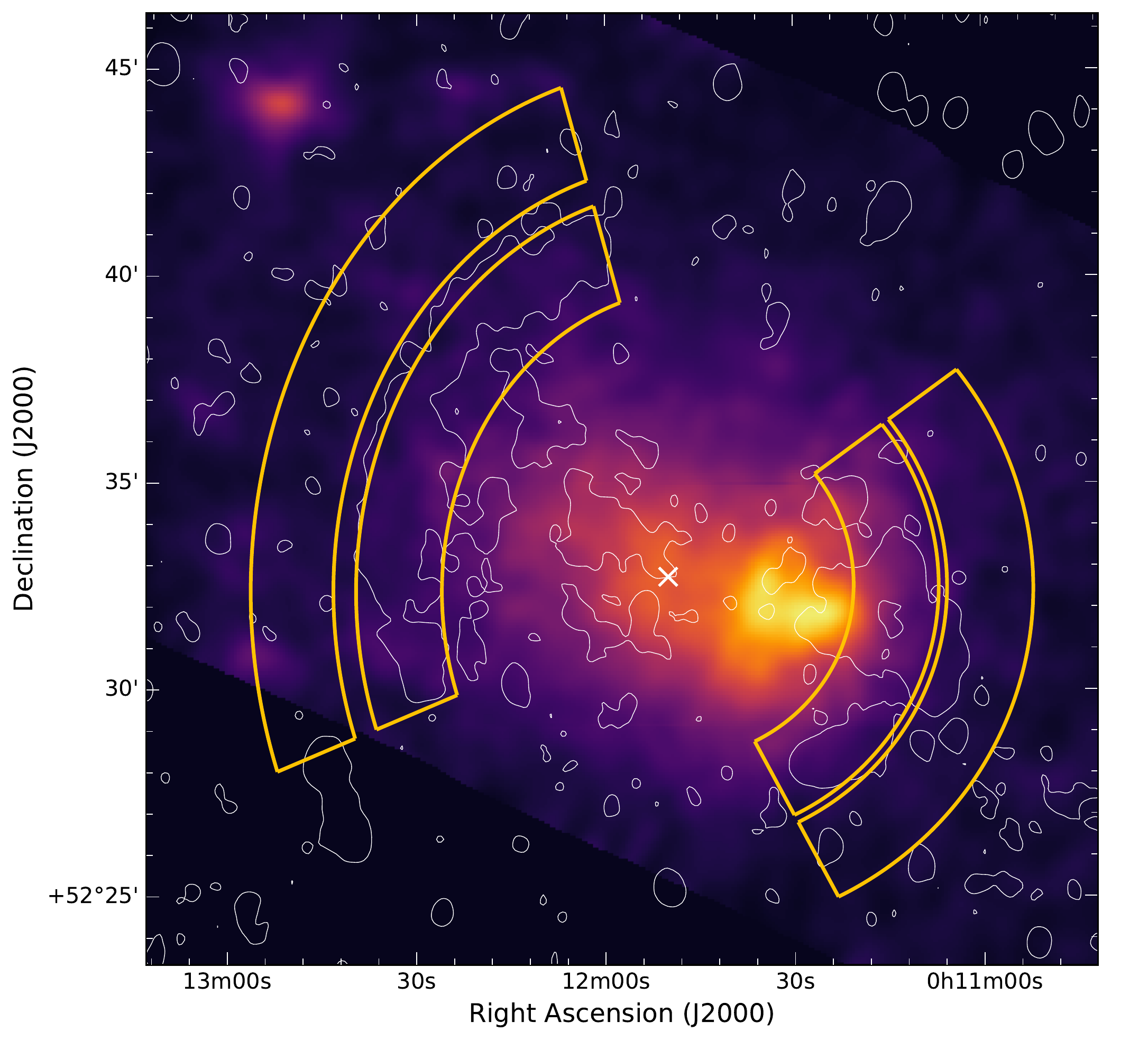}
\caption{{\it Suzaku} 0.5--4.0 keV image of ZwCl\,0008. WSRT radio contours are drawn in white at the $3\sigma_{\rm rms}$ level. Orange sectors overlaid represent the regions where the temperature measurements were extracted for the western and eastern relic (see left and right panel in Fig. \ref{fig:temp}, respectively). As for Fig. \ref{fig:SB-T_sectors}, the white cross represents the cluster center.}\label{fig:suzaku}
\end{figure}

The spatial and spectral analysis results are shown in Sects. \ref{sec:west}, \ref{sec:east} and \ref{sec:bullet} and the best-fit values reported in Tabs. \ref{tab:pyxel_res} and \ref{tab:temp_prof}. 
The corresponding MCMC ``corner plots'' for the distribution of the uncertainties in the fitted parameters of the surface brightness analysis are shown in Appendix \ref{apdx:cornerplots}. We used the distribution on the compression factor to obtain the uncertainties on $\mathcal{M}_{S_X}$, while the uncertainties on $\mathcal{M}_{T_X}$ have been calculating with 2,000 Monte Carlo realizations of Eq.~\ref{eq:mach_t}.

\subsection{The western sector}\label{sec:west}
\begin{figure*}
\centering
{\includegraphics[width=0.49\textwidth]{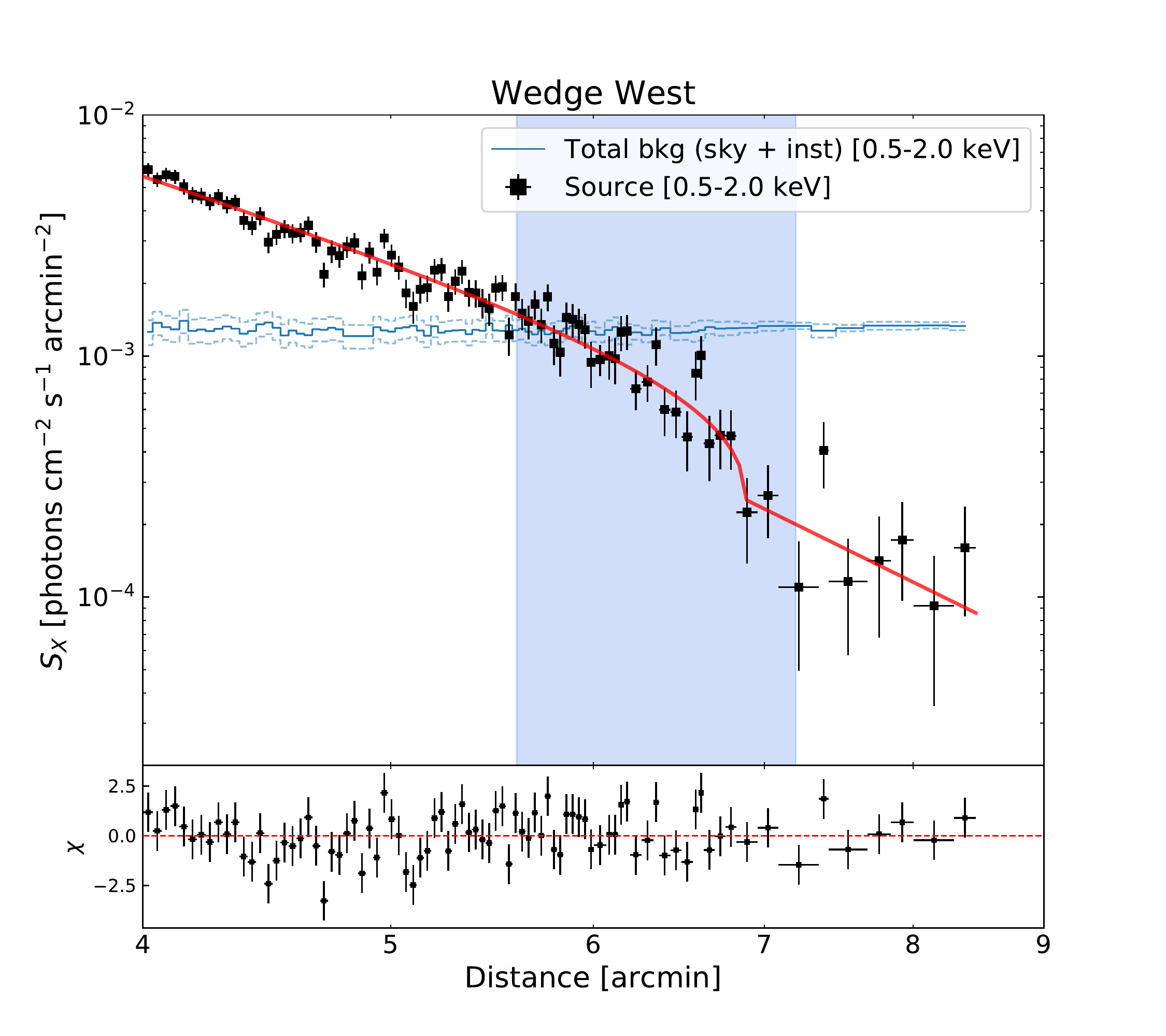}}
{\includegraphics[width=0.49\textwidth]{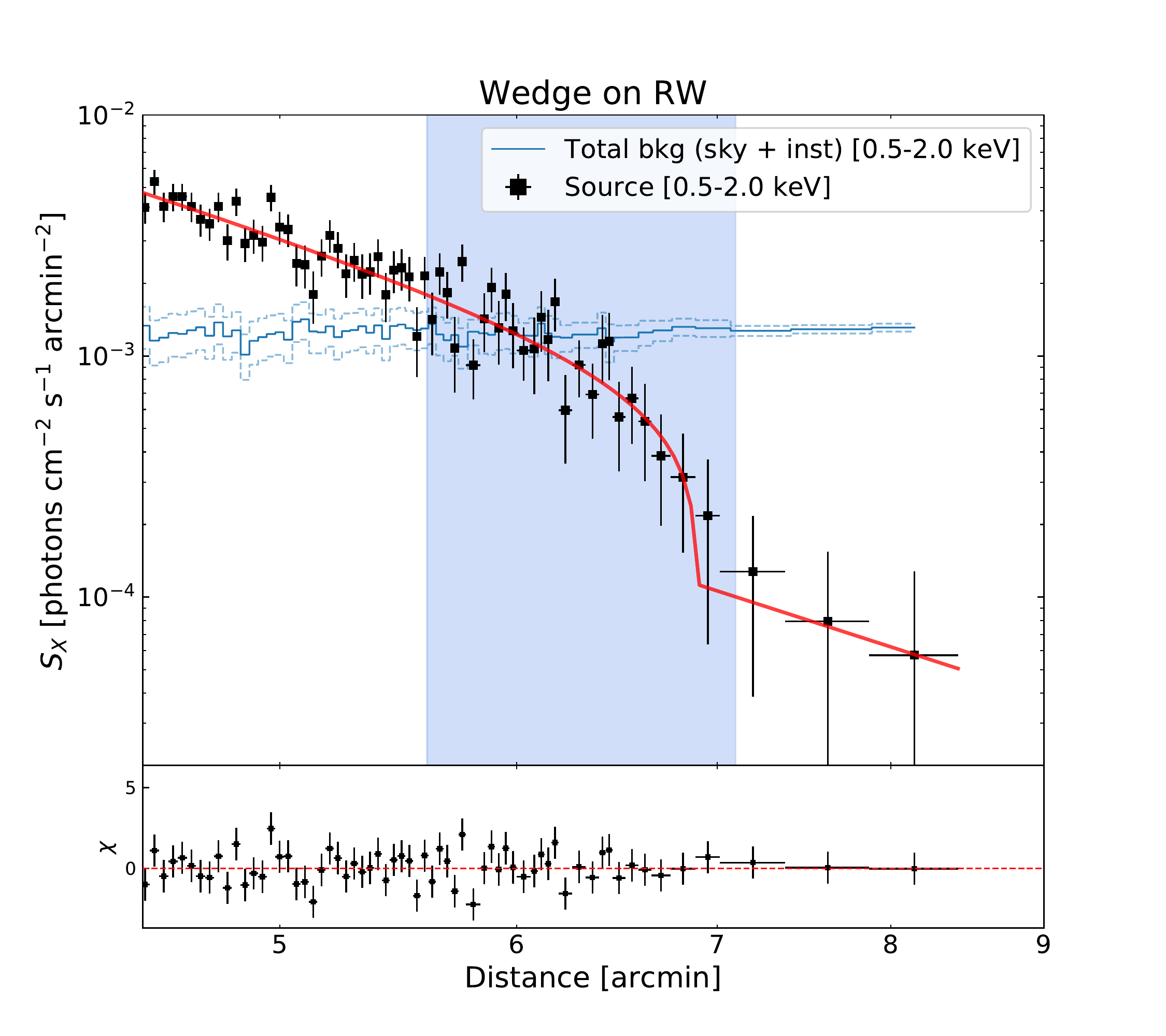}}\\
{\includegraphics[width=0.49\textwidth]{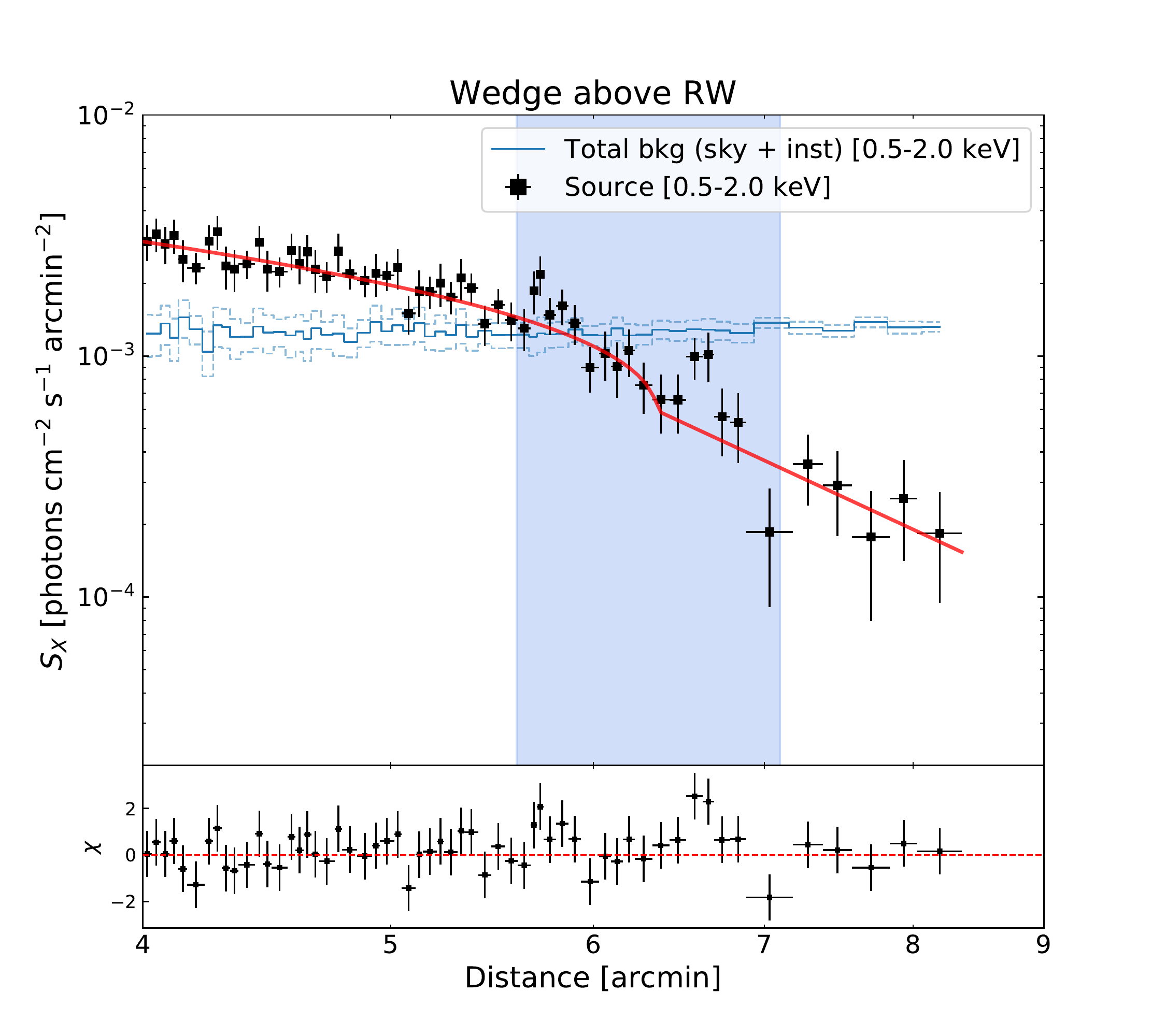}}
{\includegraphics[width=0.49\textwidth]{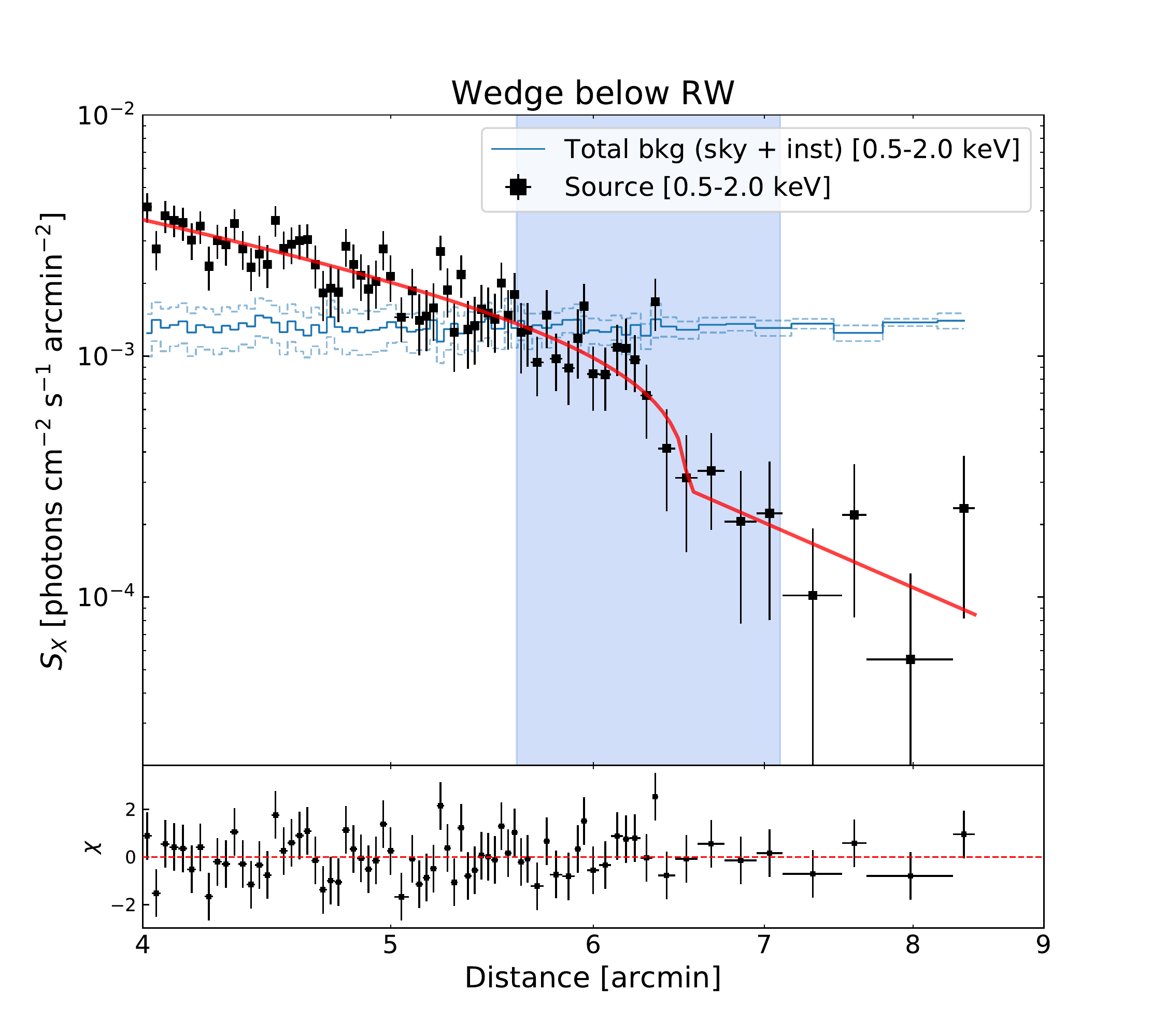}}\\
\caption{Surface brightness profiles across the western sector (top left panel) and the sub-sectors on, above and below RW (top right, bottom left and bottom right panels, respectively).The light blue rectangle identifies the position of the western radio relic. The total background level (i.e. instrumental and astrophysical) is shown by the light blue line, with the $\pm1\sigma$ uncertainties (light blue dashed lines). On the bottom of each panel, the residuals (i.e. $\frac{S_{\rm X,obs}-S_{\rm X,mod}}{\Delta S_{\rm X,obs}}$) are displayed.}\label{fig:sb_west}
\end{figure*}

The best-fitting double power-law model finds  the presence of a density jump with $\mathcal{C}=1.70_{-0.64}^{+1.04}$ located at $r=6.88_{-0.27}^{+0.15}$ arcmin (i.e. $\sim700$ kpc, at the ZwCl\,0008 redshift) from the cluster center (top left panel in Fig.~\ref{fig:sb_west}). Assuming the Rankine-Hugoniot density jump condition, this results in a Mach number for the western edge of $\mathcal{M}_{S_X}=1.48_{-0.32}^{+0.50}$ (Eq.~\ref{eq:mach_c}), which shows a shock detection at the $\sim90\%$ confidence level. No significant differences have been found by varying the background level by $\pm5\%$ (i.e., three times the residual fluctuation in the 9--12 keV band). The same region was also fitted with a simple power-law model, representative of the surface brightness profile at the cluster outskirts in the absence of shock discontinuities. We compared the results of the two models performing the Bayesian Information Criterion \citep[BIC, see][]{kass+raftery95} analysis, for which the model with the lower score is favored. We obtain BIC=195 ($\chi^2=186.35$) and the BIC=126 ($\chi^2=104.30$) for the power-law and the broken power-law model respectively, again pointing to the presence of a discontinuity at the western relic position. 

The temperature profile, derived across this discontinuity, shows the presence of heated gas behind the edge and colder gas  ahead of it ($kT_1 = 8.55^{+1.35}_{-1.14}$ and $kT_2=3.01^{+1.12}_{-0.70}$~keV, respectively, see filled blue squares in the left panel in Fig. \ref{fig:temp}). When obtained consistent result when we decrease the sector width by a factor of two (see empty blue squares in the left panel in Fig. \ref{fig:temp}). In principle, the temperature jump at the shock is also affected by the intrinsic temperature gradient of the cluster, before the shock passage \citep{vikhlinin+06}. Following \cite{burns+10}, the expected temperature variation in our temperature bin is about 0.7 keV (see solid line in the right panel in Fig \ref{fig:temp}). We add this variation as a systematic uncertainty in the temperature estimation.
Additional support for the presence of heated gas behind the detected edge is that we do not find  significant variation of temperature in the north and south directions (see the red and green sectors in the right panel in Fig. \ref{fig:SB-T_sectors} and temperature profile in the central panel in Fig. \ref{fig:temp}), where indeed there is no evidence of shocks. We also investigated possible systematic uncertainties associated with Galactic abundance ($N_{\rm H}$) variations across the cluster, using the ${\rm E(B-V)}$ reddening map at 100 $\mu$m from the NASA/IPAC Infrared Science Archive (IRSA) \footnote{\url{https://irsa.ipac.caltech.edu/cgi-bin/bgTools/nph-bgExec}} \citep{schlegel+98} and  assuming $N_{\rm H}\propto{\rm E(B-V)}$. We found a mild $N_{\rm H}$ variation (e.g. $\sim9\%$) in the west with respect to the cluster center value. The fit was then repeated, adding/subtracting this fluctuation and keeping $N_{\rm H}$ fixed, showing an increase of the temperature uncertainties of about $^{+0.9}_{-0.5}$ and $^{+0.2}_{-0.1}$ in the post- and pre-shock regions, respectively.
We use the drop in the temperature at the western edge, i.e. $\mathcal{R}=2.61^{+1.03}_{-0.69}$, to obtain the Mach number of the shock, i.e., $\mathcal{M}_{T_X}=2.35^{+0.74}_{-0.55}$ (see Eq.~\ref{eq:mach_t}).

Additional temperatures were derived in the relic sectors from the {\it Suzaku} observations (see orange sectors in Fig. \ref{fig:suzaku}). The abundance and Galactic absorption have been fixed at the same values as the {\it Chandra} observations, assuming a \texttt{phabs*apec} model and adopting the \cite{lodders+09} abundance table. The sky background was estimated using the {\it ROSAT} background tool, with the intensity of the cosmic X-ray background (CXB) allowed to change by $\pm10\%$ to explain cosmic variance. Given the high sensitivity of {\it Suzaku}, the spectra were grouped to have a minimum of 20 counts per bin, and the $\chi^2$ statistic was used. 
The temperature estimated in the post-shock region with {\it Suzaku} is $kT_1=4.67^{+1.13}_{-0.78}$, which is lower than the one obtained with {\it Chandra} at the $>90\%$ confidence level (see orange diamonds in the left panel in Fig. \ref{fig:temp}). 
We looked for possible temperature contamination from the cold front in the post-shock region, due to the limited {\it Suzaku} spatial resolution (i.e. $\sim2$ arcmin), by reducing the width of the post-shock region to $30^{\prime\prime}$: no significantly different temperature has been found. The difference in temperature in the post-shock region between {\it Chandra} and {\it Suzaku} might be explained by different instrumental calibrations. Cross-correlation studies of {\it XMM-Newton}/{\it Suzaku} \citep{kettula+13} and {\it XMM-Newton}/{\it Chandra} \citep{schellenberger+15} have shown that {\it Chandra} finds systematically higher temperatures, up to 20--25\% for cluster temperatures of 8~keV, compared with {\it XMM-Newton} \citep{schellenberger+15}. On the contrary, differences between {\it Suzaku} and {\it XMM-Newton} result to be negligible \citep{kettula+13}. On the other hand, the pre-shock temperature from {\it Suzaku} agrees well with the {\it Chandra} measurement (i.e. $kT_2=2.38^{+0.23}_{-0.21}$ and $kT_2=3.01^{+1.12}_{-0.70}$ keV, respectively), suggesting that standard blank sky field and background modelling give consistent results. Including also the systematic uncertainties (i.e. global temperature profile and instrumental calibrations), we found $\mathcal{M}_{T_X}=2.02^{+0.74}_{-0.43}$ with {\it Suzaku}, which is within the $1\sigma$ confidence level with the {\it Chandra} result.

The pressure jump across the edge is $4.45^{+2.00}_{-1.47}$.
Using the {\it Chandra} pre-shock temperature $kT_2=3.01^{+1.12}_{-0.70}$~keV and the Mach number given by the {\it Chandra} temperature profile, we obtain a shock velocity of $v_{\rm shock,W}=1989^{+509}_{-468}$~km~s$^{-1}$. Given the distance of the edge from the cluster center ($\sim7$ arcmin, i.e. $\sim780$~kpc) and the shock velocity, we estimated the time since the first core passage being $\sim0.3-0.5$~Gyr, older than the time found for the Bullet Cluster \cite{markevitch06} and for Abell\,2146 \citep{russell+10}, i.e. $\sim0.2$ Gyr. The time we found is consistent with the one found by \cite{golovich+17} assuming an ``outbound'' scenario, i.e. $0.49-1.0$~Gyr.

The most remarkable aspect of ZwCl\,0008 is that the western radio relic traces only part of the shock front ($\rm LLS_{RW}\approx290~kpc$, while $\rm LLS_{edge,W}\approx1~Mpc$). A possible explanation is that the Mach number of the shock varies along the length of the edge and the relic forms only where $\mathcal{M}$ is high enough to accelerate electrons. To investigate this, we divided the western edge into three sub-sectors, tracing the shock above, below, and on RW (see left panel in Fig.~\ref{fig:SB-T_sectors} and Table~\ref{tab:pyxel_res}). The corresponding surface brightness profiles are displayed in the top right, bottom right, and bottom left panels in Fig.~\ref{fig:sb_west}. Due to the low S/N in the upstream region, for these sectors we additionally constrained the slope $a_2$ to be in the range $1<a_2<3.2$. Those values have been chosen to match the slopes of the surface brightness profiles, at $R_{500}$, 
of the full cluster sample in the {\it Chandra}--{\it Planck} Legacy Program for Massive Clusters of Galaxies\footnote{\url{hea-www.cfa.harvard.edu/CHANDRA\_PLANCK\_CLUSTERS/}} \citep[PI: C. Jones;][Andrade-Santos et al., in prep.]{andrade-santos+17}. 
Under these assumptions, we obtain $\mathcal{M}_{\rm RW}^{\rm above}=1.30_{-0.17}^{+0.46}$, $\mathcal{M}_{\rm RW}^{\rm on}=2.98_{-0.85}^{+2.62}$ and $\mathcal{M}_{\rm  RW}^{\rm below}=1.70_{-0.55}^{+0.79}$ for the sub-sector above, on, and below the western relic, respectively. They are consistent to each other within the error bars, hence we cannot assert whether the Mach number is varying along the western X-ray discontinuity. Given the few counts in the pre- and post-shock regions, we were not able to perform a temperature analysis for the  three separate sub-sectors.

\subsection{The eastern sector}\label{sec:east}
\begin{figure}
\centering
\includegraphics[width=0.5\textwidth]{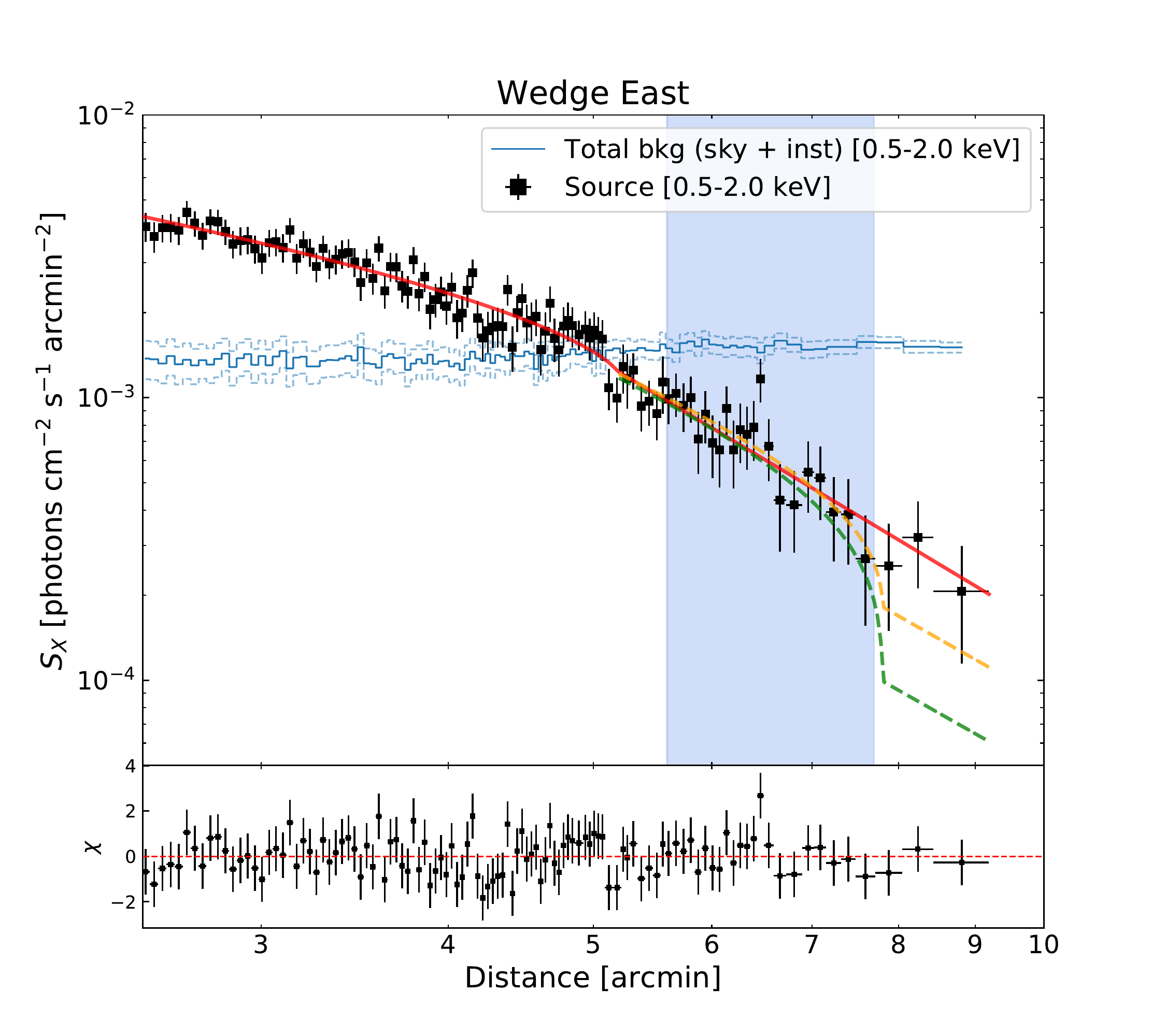}
\caption{Surface brightness profile across the eastern sector. The light blue rectangle identifies the position of the eastern radio relic.
The total background level (i.e. instrumental and astrophysical) is shown by the light blue line, with the $\pm1\sigma$ uncertainties (light blue dashed lines). On the bottom, the residuals (i.e. $\frac{S_{\rm X,obs}-S_{\rm X,mod}}{\Delta S_{\rm X,obs}}$) are displayed, representative of the broken power-low best-fit (red line). Models of density jumps of $\mathcal{C}=1.7$ and $\mathcal{C}=2.3$ at fixed $r_{\rm break}=7.8$ arcmin are also overlaid (dashed orange and green lined, respectively).}\label{fig:sb_east}
\end{figure}

No clear discontinuity is detected in the east. Assuming the broken power-law model, as suggested by the presence of the radio relic (RE), we found a mild jump in density (Fig.~\ref{fig:sb_east}) of $\mathcal{C}=1.09_{-0.08}^{+0.11}$ at $5.16_{-0.23}^{+0.26}$ arcmin (i.e. $\sim 550$ kpc from the cluster center), suggesting simply a change of slope at this location \citep[i.e. a King profile, see][]{king72}. However, BIC scores slightly disfavor a $\beta$-model  \citep[see][]{cavaliere+fusco-femiano76}, rather than the broken power-law model (BIC=108 against BIC=100, respectively). Interestingly, the location of this putative X-ray discontinuity is displaced from the edge of the eastern relic (i.e. $r\sim7.8$ arcmin)
toward the cluster center. No drop has been detected at the relic location, either from the X-ray image and surface brightness profiles (Figs.~\ref{fig:SB-T_sectors} and \ref{fig:sb_east}). However, we note that this relic is located far from the cluster center, i.e. $\sim5.6-7.8$ arcmin, or $\sim610-900$ kpc, at the edge of the field of view (FOV) of our observation (see the right panel in Fig. 1). Hence, not all the ObsIDs cover the area ahead the eastern relic, i.e. the pre-shock region.
In Fig. \ref{fig:sb_east} we also overlay models of a density jump of $\mathcal{C}=1.7$ (i.e. $\mathcal{M}=1.5$, see orange dashed line) and $\mathcal{C}=2.3$ (i.e. $\mathcal{M}=2.0$, see green dashed line), in the region $5\lesssim r \lesssim 9$ arcmin\footnote{In this way, we avoid the change of slope at $r\sim5$ arcmin.}, with $r_{\rm break}$ fixed at the outermost edge of the eastern relic (i.e. $r_{\rm RE}=7.8$ arcmin). It is clear that a density jump of $\mathcal{C}=2.3$ is ruled out by our data. On the other hand, a density jump of $\mathcal{C}=1.7$ is still consistent with our observations.
Hence, we conclude that, if present, a shock front at the location of the eastern relic should be quite weak (i.e. $\mathcal{M}\lesssim1.5$). In agreement with this result, we obtain a temperature based Mach number from {\it Suzaku} of $\mathcal{M}=1.54^{+0.65}_{-0.47}$ at the relic position (see orange sectors Fig. \ref{fig:suzaku}).

\begin{figure}
\centering
\includegraphics[width=0.5\textwidth]{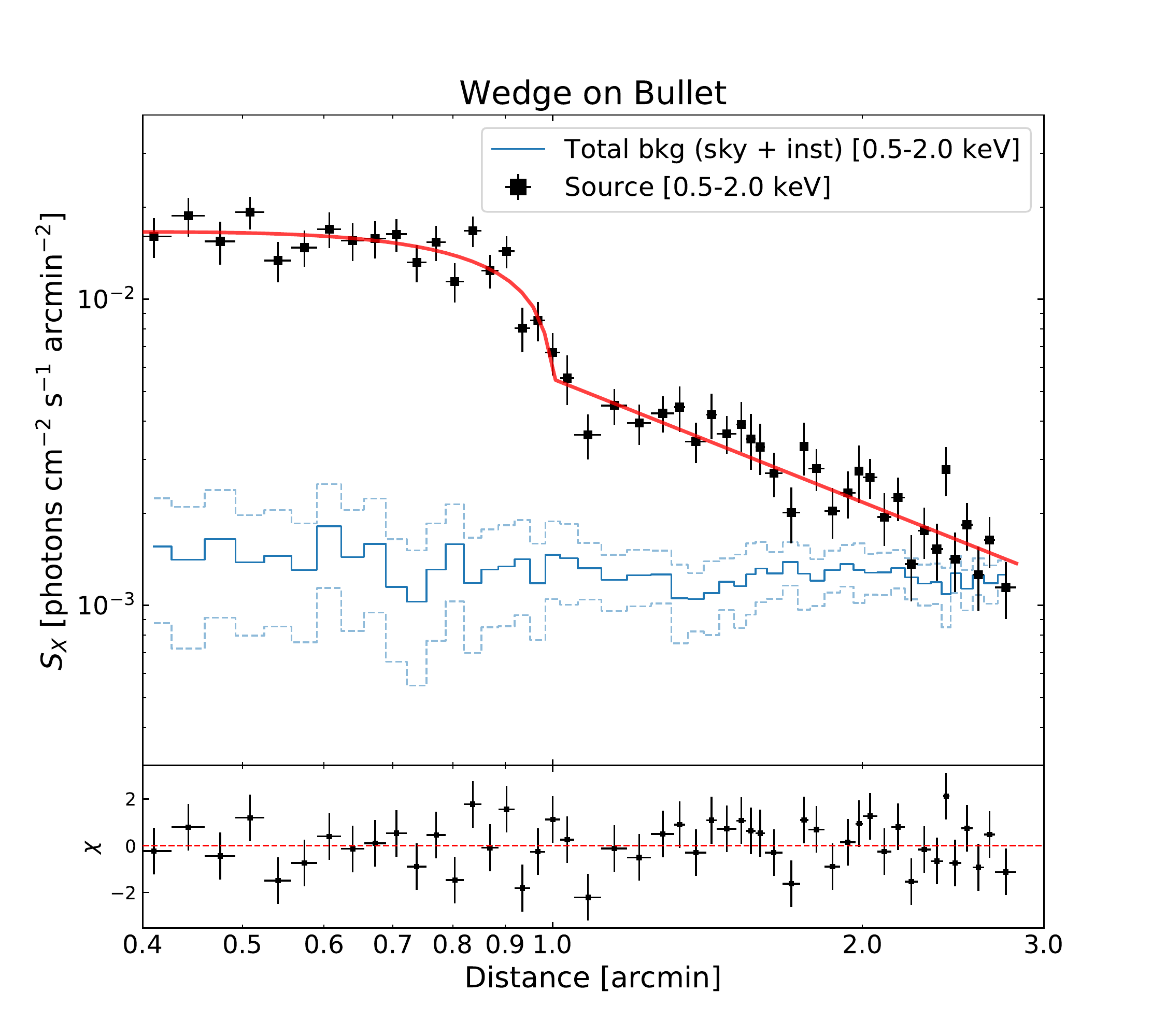}
\caption{Surface brightness profile across the bullet sector. The total background level (i.e. instrumental and astrophysical) is shown by the light blue line, with the $\pm1\sigma$ uncertainties (light blue dashed lines). On the bottom, the residuals (i.e. $\frac{S_{\rm X,obs}-S_{\rm X,mod}}{\Delta S_{\rm X,obs}}$) are displayed.}\label{fig:sb_cf}
\end{figure}

\begin{figure*}
\centering
\hspace{-4mm}
{\includegraphics[width=0.35\textwidth]{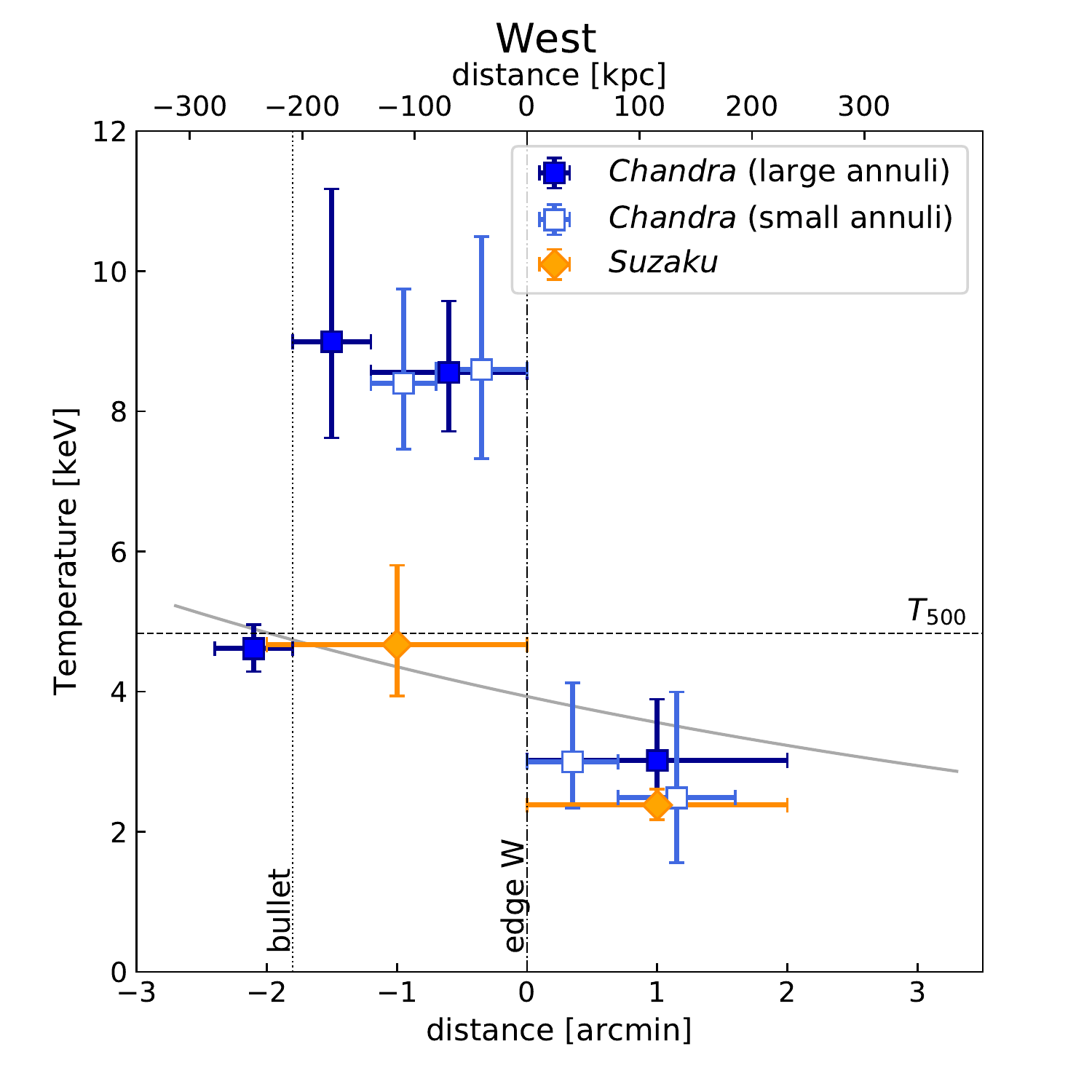}}
\hspace{-5mm}
{\includegraphics[width=0.35\textwidth]{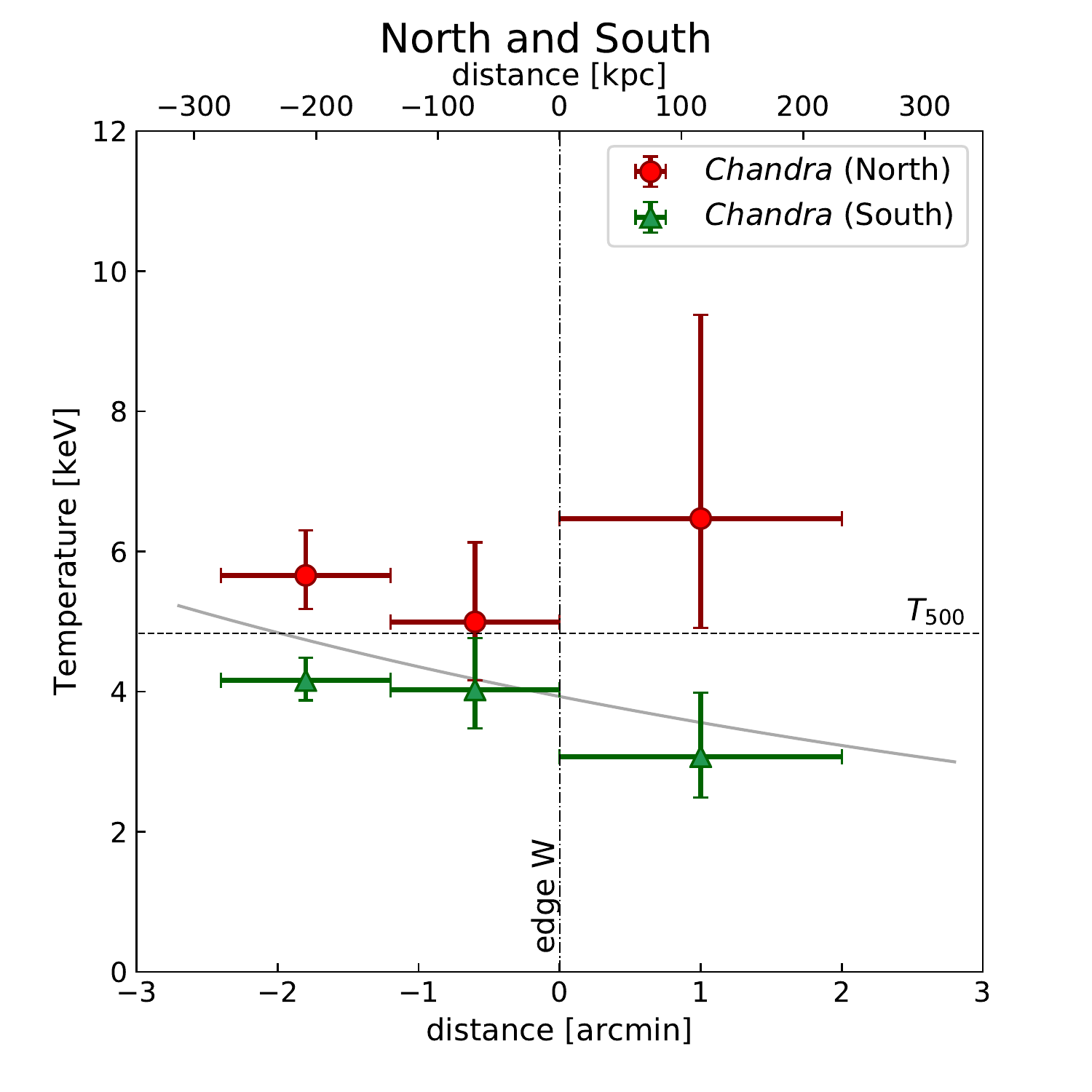}}
\hspace{-5mm}
{\includegraphics[width=0.35\textwidth]{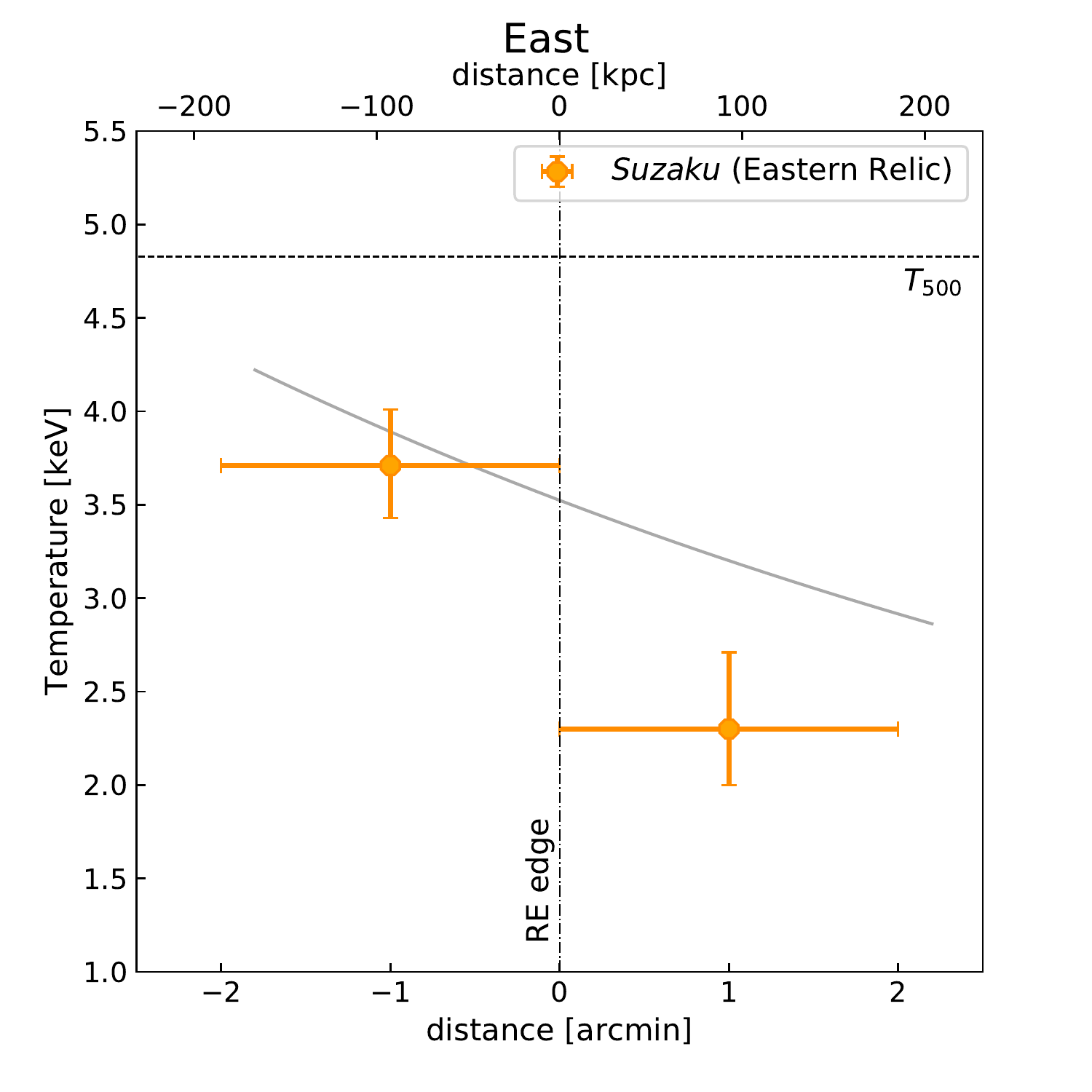}}
\caption{Radial temperature profiles westward (left), northward and southward (central) and on the eastern relic (right). All the values have been obtained by fixing the abundance and hydrogen column density at $A=0.3$~Z$_\odot$ and $N_{\rm H}=0.311\times10^{22}$~cm$^2$, respectively.
The horizontal dashed lines in the three panels represent the averaged temperature of the global cluster at $R_{500}$, obtained with $\it Chandra$. The vertical dot-dashed lines display the position of the western edge (left and central panel) and the edge of the eastern relic (right panel); the vertical dotted line in the left panel displays the position of the cold front. The solid gray line represents the averaged temperature profile according to \cite{burns+10}.}\label{fig:temp}
\end{figure*}

\subsection{The bullet sector}\label{sec:bullet}
In order to match the curvature of the bullet, we chose  
an elliptical sector displaced from the cluster center by $\sim3.5^\prime$ (${\rm RA=0^h11^m25^s.976}$ and ${\rm DEC=+52^\circ31^\prime58^{\prime\prime}.49}$, J2000). 
The best-fit of the surface brightness profile analysis (Fig.~\ref{fig:sb_cf}) results in a density jump $\mathcal{C}=2.06_{-0.19}^{+0.24}$ 
at $r=0.99\pm0.02$ arcmin from the sector center (i.e. $\sim490$ kpc from the cluster center, at the cluster redshift). 
At this location, we measure a temperature jump of $T_1/T_2=0.56^{+0.10}_{-0.09}$ (see Fig.~\ref{fig:temp}). 
By combining the temperature and the electron density jumps, we obtain 
$P_1/P_2=1.15^{+0.25}_{-0.20}$, consistent with a constant pressure across the edge, confirming that the discontinuity is a cold front.

\section{Discussion}\label{sec:disc}
At the location of a shock front, particles are thought to be accelerated via first-order Fermi acceleration, e.g. {\it diffusive shock acceleration} \citep[DSA,][]{drury83,blandford+eichler87} and {\it shock drift acceleration} \citep[SDA,][]{wu84,krauss-varban+wu89} mechanisms.
In particular, the SDA process has been recently invoked to solve the so-called ``electron injection problem'', which is particularly important in the low-$\mathcal{M}$ regime (i.e., $\mathcal{M}\lesssim2$), giving the necessary pre-acceleration to the electron population to facilitate the DSA process \citep{guo+14a,guo+14b,caprioli+spitkovsky14}. 
The interaction between these accelerated particles and the amplified magnetic field in merging clusters produces synchrotron emission in the form of radio relics. According to the DSA theory, there is a relation between the spectral index measured at the shock location, the so-called injection spectral index $\alpha_{\rm inj}$, and the Mach number $\mathcal{M}$ of the shock \citep[e.g.][]{giacintucci+08}:

\begin{equation}\label{eq:mach_radio}
\mathcal{M}_{\rm radio} = \sqrt{\frac{2\alpha_{\rm inj} + 3}{2\alpha_{\rm inj} -1}} \, .
\end{equation}

Thus for DSA, the Mach number estimated in this way is expected to agree with the one obtained from the X-ray observations. This is not always the case: a number of radio relics have been found to have higher radio Mach numbers than the one obtained via X-ray observations \citep[e.g.][]{macario+11,vanweeren+16,pearce+17}. Another problem is that in some cases no radio relics have been found even in the presence of clear X-ray discontinuities \citep[e.g.][]{shimwell+14}. Furthermore, it is still unclear whether the DSA mechanism of thermal electrons, in case of low-$\mathcal{M}$ shocks, can efficiently accelerate particles to justify the presence of giant radio relic \citep[e.g.][]{brunetti+jones14,vazza+14,vanweeren+16,hoang+17}.

Several arguments have been proposed to address the issues described above.
One possibility is that the assumption of spherical symmetry, which is at the basis of Eq.~\ref{eq:mach_c} and \ref{eq:mach_t}, is not strictly correct, and that projection effects can hide the surface brightness and temperature discontinuity, leading to smaller $\mathcal{M}$ from the X-ray compared to the one obtained from the radio analysis. 
Also, the Mach number might be not constant across the shock front, as it is suggested by numerical simulations \citep[e.g.][]{skillman+13}, and synchrotron emission is biased to the measurement of high Mach number shocks \citep{hoeft+bruggen07}. 
An alternative explanation is given by invoking the {\it re-acceleration} mechanism \citep[e.g.][]{markevitch+05,macario+11,bonafede+14,shimwell+15,botteon+16b,kang+17,vanweeren+17a}. Indeed, several very recent observations \citep{vanweeren+17a,vanweeren+17b,degasperin+17,digennaro+18} revealed that if a shock wave passes through fossil (i.e. already accelerated) plasma, such as the lobes of a radio galaxy, it could re-accelerate or re-energize the electrons and produce diffuse radio emission. 

In order to best investigate the properties of shocks in ZwCl\,0008, in the following sections we will discuss the comparison between our new {\it Chandra} observations and the previous radio analysis by \cite{vanweeren+11}.

\subsection{Radio/X-ray comparison for the western relic}
The previous radio analysis of ZwCl\,0008 was performed at 241, 610, 1328 and 1714~MHz with the GMRT and the WSRT \citep{vanweeren+11}. This work revealed the presence of two symmetrically located radio relics (see also right panel of Fig. \ref{fig:chandra}). In the proximity of the western relic our {\it Chandra} observations indicate the presence of a shock. From the spectral index analysis\footnote{$\alpha_{\rm inj}$ was calculated either directly from the map, and from the volume-integrated spectral index $\alpha_{\rm int}$ \citep[i.e. $\alpha_{\rm inj}= \alpha_{\rm int}+0.5$,][]{blandford+eichler87}. The two values are consistent with each other.} 
of RW, \citeauthor{vanweeren+11} estimated $\alpha_{\rm inj}=-1.0\pm0.15$, 
with a spectral index steepening towards the cluster center (i.e. in the shock downstream region) due to synchrotron and Inverse Compton energy losses, as expected from an edge-on merger event \citep[see Fig.~8 in][]{vanweeren+11}. 
Given the injection spectral indices and Eq.~\ref{eq:mach_radio}, \citeauthor{vanweeren+11} estimated a radio Mach numbers of $\mathcal{M}_{\rm RW}=2.4^{+0.4}_{-0.2}$. This value is consistent within the uncertainties with our X-ray analysis ($\mathcal{M}_{S_X}=1.48^{+0.50}_{-0.32}$ and $\mathcal{M}_{T_X}=2.35^{+0.74}_{-0.55}$), consistent with the DSA scenario for the western relic's origin.

An interesting complication to this picture comes by the fact that the western relic only partly traces the shock front. Total or partial absence of relic emission in presence of clear X-ray discontinuities could be explained by having a shock strength which drops below a certain threshold, depending on the plasma beta parameter ($\beta\equiv P_{\rm gas}/P_{B}$) at the shock \citep{guo+14a,guo+14b}. 
Unfortunately, the net count statistics in those sectors are very poor and our estimated Mach numbers in the three sub-sectors are characterized by large error bars (see Table~\ref{tab:temp_prof}). Hence, we cannot assert whether $\mathcal{M}$ variations are present and justify the smaller size of RW compared to the X-ray shock extent (however, see Sect. \ref{sec:efficiency}).
Another appealing explanation for the origin of the western relic is suggested by the proximity of three different radio galaxies (i.e. sources C, E and F in the right panel in Fig.~\ref{fig:chandra}) which can provide the fossil electrons for the synchrotron emission, according to the re-acceleration mechanism. In this case, the absence of diffuse radio emission associated to the relic, above and below RW, can be simply explained by the absence of underlying fossil plasma to be re-accelerated by the crossing shock wave. 
For the case of ZwCl\,0008, there is no clear connection between the radio galaxies and RW, which is the strongest requirement to invoke the re-acceleration mechanism, together with the detection of the shock. However, such fossil plasma can be  faint and characterized by a very steep spectral index, meaning that it is best detected with sensitive low- frequency observations.

\begin{table*}
\caption{Wedges information (columns 1 to 4) and best-fit parameters (columns 5 to 8) from the surface brightness profiles shown in Figures \ref{fig:sb_west}, \ref{fig:sb_east} and \ref{fig:sb_cf}. A broken power-law model has been assumed (see Eq.~\ref{eq:mach_c}) for each sector.}
\begin{center}
\begin{tabular}{lccccccc}
\hline
\hline
Sector & $\Delta\theta$ & Min. count per bin & $e$ & $\alpha_1$ & $\alpha_2$ & $r_{\rm edge}$ & $\mathcal{C}$ \\
&[degree] & &	&		& & [arcmin]		& \\
\hline
West & 98 & 70 & 1.14 & $2.20_{-0.10}^{+0.10}$ & $3.11_{-1.75}^{+2.40}$ & $6.88_{-0.26}^{+0.15}$ & $1.70_{-0.65}^{+0.91}$ \\
above RW$^\dagger$ & 36 & 50 & 1.14 & $0.91_{-0.56}^{+0.37}$ & $2.96_{-1.35}^{+0.22}$ & $6.35_{-0.47}^{+0.66}$ & $1.44_{-0.36}^{+0.97}$ \\
on RW$^\dagger$ & 30 & 25 & 1.14 & $2.38_{-0.25}^{+0.25}$ & $2.50_{-1.09}^{+0.64}$ & $6.89_{-0.16}^{+0.17}$ & $2.99_{-0.86}^{+0.90}$ \\
below RW$^\dagger$ &  32 & 30 & 1.14 & $1.44_{-0.33}^{+0.27}$ & $2.82_{-1.37}^{+0.37}$ & $6.53_{-1.09}^{+0.25}$ & $1.96_{-0.94}^{+1.30}$ \\
East$^\ddagger$ & 107 & 70 & 1.14 & -- & -- & 7.8 & $\lesssim 1.7$ \\
Bullet & 60 & 40 & 1.42 & $-0.24_{-0.29}^{+0.23}$ & $1.17_{-0.07}^{+0.06}$ & $0.99_{-0.02}^{+0.02}$ & $2.06_{-0.19}^{+0.24}$ \\
\hline
\end{tabular}
\end{center}
{Note: All the sectors are centered in the cluster center (i.e. ${\rm RA=0^h11^m50^s.024}$ and ${\rm DEC=+52^\circ32^\prime37^{\prime\prime}.98}$, J2000), 
 with the exception of the bullet (${\rm RA=0^h11^m25^s.976}$ and ${\rm DEC=+52^\circ31^\prime58^{\prime\prime}.49}$, J2000). The ellipticity of each sector is given by the parameter $e$.  $^\dagger$Prior on $a_2$ (see Sect. \ref{sec:west}) $^\ddagger$ Model.} 
\label{tab:pyxel_res}
\end{table*}

\begin{table*}
\caption{Best-fit temperature profiles for the X-ray discontinuities. A \texttt{phabs*APEC} model with fixed $N_{\rm H}=0.311\times10^{22}$ cm$^{-2}$ and $A=0.3$ Z$_\odot$ has been assumed for the analysis.}
\begin{center}
\begin{tabular}{lcccccccc}
\hline
\hline
Sector & Instrument & \multicolumn{2}{c}{$kT$}  & \multicolumn{2}{c}{$^{\rm stat}/_{\rm dof}$} & $\mathcal{R}$ & $\mathcal{M}_{T_X}$ & $\mathcal{M}_{S_X}^\diamond$ \\
       & & \multicolumn{2}{c}{[keV]} & & \\
\hline
$R_{500}$ & {\it Chandra} & \multicolumn{2}{c}{$4.83\pm0.06$} &  \multicolumn{2}{c}{$^{4214.75}/_{3785}$} & -- & -- & -- \\
\multirow{2}{*}{West} & {\it Chandra} &  $8.55^{+1.35\,(a)}_{-1.14}$ & $3.01^{+1.12\,(b)}_{-0.70}$ & $^{3643.65}/_{3964}$$^{(a)}$ & $^{1182.10}/_{1258}$ $^{(b)}$ & $2.61^{+1.03}_{-0.69}$ & $2.35^{+0.74}_{-0.55}$ & $1.48_{-0.32}^{+0.50}$ \\
& {\it Suzaku} & $4.67^{+1.13\,(a)}_{-0.78}$ & $2.38^{+0.23\,(b)}_{-0.21}$ & $^{47.39}/_{54}$$^{(a)}$ & $^{208.61}/_{228}$$^{(b)}$& $2.05^{+0.77}_{-0.43}$ & $2.02^{+0.74}_{-0.43}$ & -- \\
above RW & {\it Chandra} & -- & -- & -- & -- & -- & -- & $1.30_{-0.17}^{+0.46}$ \\
on RW   & {\it Chandra} & -- & -- & -- & -- & -- & -- & $2.98_{-0.85}^{+2.62}$ \\
below RW & {\it Chandra} & -- & -- & -- & -- & -- & -- & $1.70_{-0.55}^{+0.79}$ \\
East & {\it Chandra} & -- & -- & -- & -- & -- & -- & $\lesssim 1.5^\ddagger$\\
on RE   & {\it Suzaku} & $3.71^{+0.30}_{-0.28}$$^{(a)}$ & $2.30^{+0.41}_{-0.30}$$^{(b)}$ & $^{309.86}/_{337}$$^{(a)}$ & $^{171.82}/_{162}$$^{(b)}$ & $1.54^{+0.39}_{-0.26}$ & $1.54^{+0.65}_{-0.47}$ & -- \\
Bullet & {\it Chandra} & $4.61^{+0.34\,(a)}_{-0.33}$ & $8.99^{+2.17\,(b)}_{-1.37}$ & $^{1250.23}/_{1602}$$^{(a)}$ & $^{1059.42}/_{1309}$ $^{(b)}$ & $0.56^{+0.10}_{-0.09}$ & -- & --\\
\hline
\end{tabular}
\end{center}
{Note: values at $^{(a)}$  $r \leq r_{\rm edge}$ and $^{(b)}$  $r > r_{\rm edge}$; $^\diamond$ calculated from $\mathcal{C}$ in Table~\ref{tab:pyxel_res}. $^\ddagger$ Model. The uncertainties on $\mathcal{M}_{S_X}$ have been obtained from the compress factor distributions shown in Appendix \ref{apdx:cornerplots}, while the uncertainties on $\mathcal{M}_{T_X}$ have been calculated with 2,000 Monte Carlo realizations of Eq.~\ref{eq:mach_t} and including the systematic uncertainty given by the cluster temperature average profile (i.e. 0.7 keV). }
\label{tab:temp_prof}
\end{table*}

\subsection{The puzzle of the eastern radio relic}
Similarly to RW, the eastern relic also displays spectral steepening towards the cluster center \citep[see Fig.~8 in][]{vanweeren+11}. The measured injection spectral index is $\alpha_{\rm inj}=-1.2\pm0.2$, which corresponds to a Mach number of $\mathcal{M}_{\rm}=2.2^{+0.2}_{-0.1}$, under the assumption of DSA of thermal electrons \citep[Eq.~\ref{eq:mach_radio} and][]{vanweeren+11b}. A surface brightness discontinuity is therefore expected in the eastward outskirts of ZwCl\,0008, tracing the shape of RE. Nonetheless, no discontinuity has been detected at the relic position in our {\it Chandra} observations.

A complication that should be taken into account is projection effects, which can hide, or at least smooth, X-ray discontinuities. Polarization analysis \citep{golovich+17} and numerical simulations \citep{kang+12} of the eastern relic showed that the merger angle in ZwCl\,0008 ranges between $25$ and $30^\circ$, being $0^\circ$ the angle associated to a perfectly edge-on collision. This possible non-negligible inclination angle might, in principle, contribute in hiding X-ray discontinuities. Despite that, our observations suggest that, if present, the shock front in the eastern side of the cluster is rather weak, i.e. $\mathcal{M}\lesssim 1.5$, which is lower than the one found by the radio spectral index analysis. Further studies, focused on this side of the cluster, are necessary to give better constraints on the strength of the putative shock front.

\subsection{Shock location and comparison with numerical simulations}
The distribution of the ICM and the exact location of the shock fronts are essential to put constraints on the characterization of 
the dynamical model of the merger event. Two previous studies have been performed for ZwCl\,0008, using weak lensing \citep{golovich+17} and N-body/hydrodynamical \citep{molnar+broadhurst17} simulations. 
Despite qualitative agreements (e.g. the identification of the most massive sub-cluster, the small impact parameter and offset of the main cluster from the dark matter peak), different sub-cluster mass ratio and time after the first core passage have been found in two works.
It is worthy to note, though, that analysis performed by \cite{molnar+broadhurst17} was based on the position of the putative shock fronts, given by the previous shallow (42 ks) X-ray observations. These were supposed to be located, in the east, at the position of the well-defined radio relic, and, in the west, further in the cluster outskirts \citep[see Fig. 1 in][]{molnar+broadhurst17}. Such positions led to an extremely high shock velocities (i.e $\sim4000$ and $5000$ km s$^{-1}$, respectively for the western and eastern shock).
This interpretation, however, does not agree with our new, deeper (410 ks), X-ray observations. We indeed detect a shock front at the western relic position, while no clear confirmation has been found at the eastern relic one (see right panel in Fig. \ref{fig:chandra}, top left panel in Fig. \ref{fig:sb_west} and Fig. \ref{fig:sb_east}). We can then conclude that, in cases of merging clusters with the presence of radio relics, the position of shock discontinuity cannot be arbitrary, but needs to match the position of the radio source. This information is particularly suitable for double radio relics, which describe merger events very close to the plane of the sky.

\subsection{Shock acceleration efficiency}\label{sec:efficiency}
As described above, one of the open questions related to the DSA mechanism is whether the particles from the thermal pool can be efficiently accelerated by a low-$\mathcal{M}$ shock (e.g. $\mathcal{M}\lesssim2$). 

The acceleration efficiency, $\eta$, is defined as the amount of kinetic energy flux available at the shock that is converted into the supra-thermal and relativistic electrons, and it relates to the synchrotron luminosity $L_{\rm sync}$ of the radio relic according to \citep{brunetti+jones14}:

\begin{equation}
\eta = \left [ \frac{1}{2} \rho_{\rm 2} v_{\rm shock}^3 \left ( 1 - \frac{1}{\mathcal{C}^2} \right ) \frac{B^2}{B^2 + B_{\rm CMB}^2} S \right ]^{-1} \Psi(\mathcal{M}) L_{\rm sync} \, ,
\end{equation}
where $\rho_{\rm 2}$ is the total density in the up-stream region, $v_{\rm shock}$ the shock speed, $\mathcal{C}$ the compression factor at the shock, $B$ the magnetic field, $B_{\rm CMB}=3.25(1+z)^2~\mu$G the magnetic field equivalent for the Cosmic Microwave Background radiation, 
and $S$ the shock surface area. Here, $\Psi(\mathcal{M})$ is a dimensionless function which takes the ratio of the energy flux injected in ``all'' the particles and those visible in the radio band \citep[see Eq.~5 in][for the exact mathematical description of $\Psi(\mathcal{M})$]{botteon+16} into account.

\begin{figure}
\centering
\hspace{-6mm}
\includegraphics[width=0.5\textwidth]{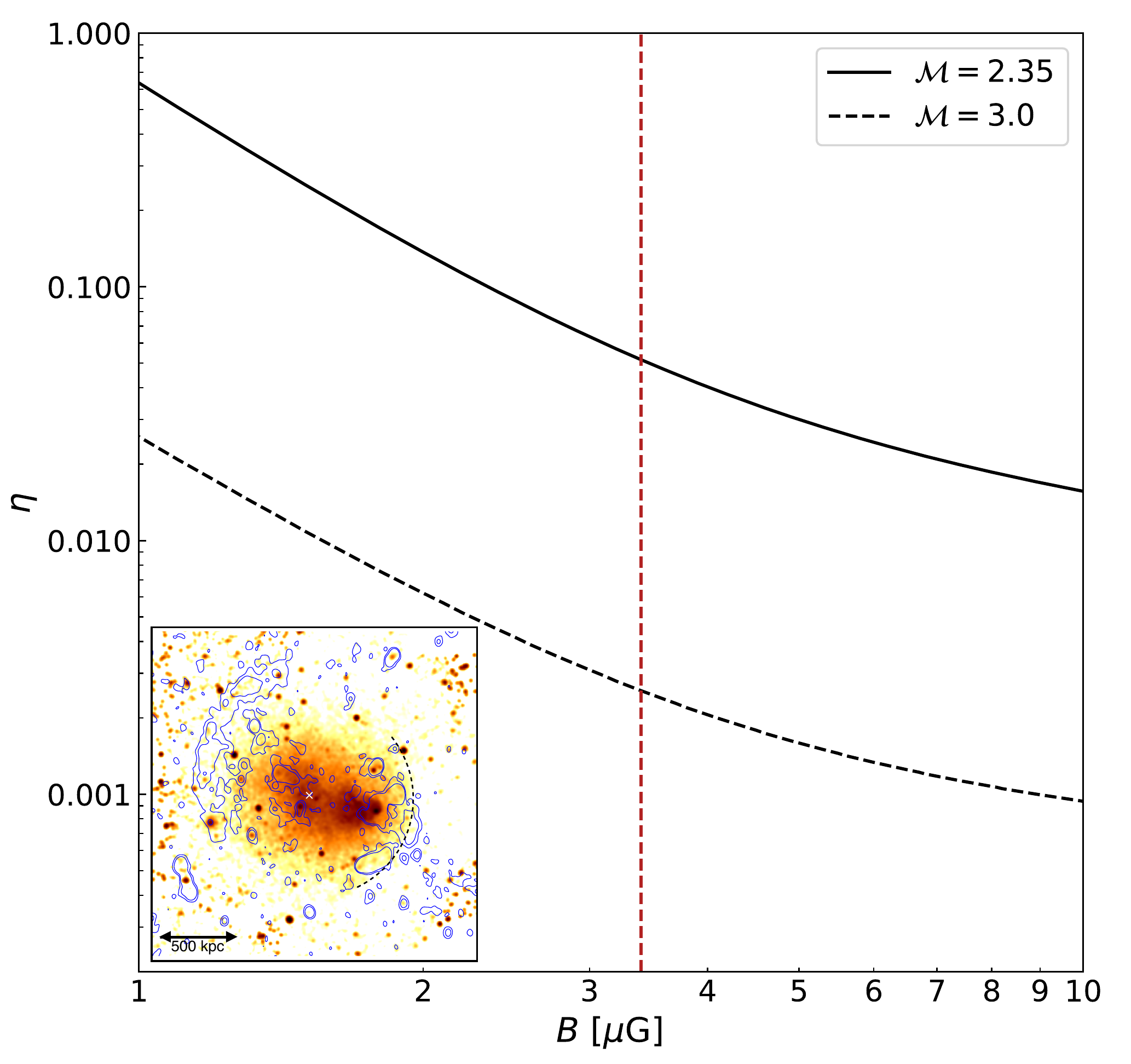}
\caption{Electron acceleration efficiency as a function of magnetic field for the western relic. The vertical red dashed line shows the value of the magnetic field estimated by \cite{vanweeren+11}. The dashed arc in the inset in the bottom left corner shows the position of the shock as revealed by the surface brightness analysis (top left panel in Fig. \ref{fig:sb_west}).}\label{fig:eff}
\end{figure}

In Fig. \ref{fig:eff} we report the electron acceleration efficiency analysis for the western radio relic, for which we have the strongest evidence of the X-ray shock, as a function of the magnetic field. We assume $S=\pi\times290^2$ kpc$^2$, $P_{\rm 1.4~GHz}=0.37\times10^{24}$  W Hz$^{-1}$ \citep[see][]{vanweeren+11}, a total pre-shock numerical density\footnote{$\rho=\mu m_{\rm H} n$} $n_{\rm 2}=1.8\times10^{-4}$~cm$^{-3}$, and a shock Mach number of $\mathcal{M}=2.35$, according to the {\it Chandra} measurement. 
Given the estimation of magnetic field of $3.4~\mu$G \citep[under the assumption of equipartition, see][]{vanweeren+11}, the efficiency required for the electron acceleration due to the shock is $\eta\sim0.05$. This would disfavor the standard DSA scenario, since efficiencies $\lesssim10^{-3}$ are expected for weak shocks \citep[e.g.][]{brunetti+jones14,caprioli+spitkovsky14,hong+14,ha+18}. 
Given the high uncertainties on our Mach number estimation, we also repeated the analysis assuming $\mathcal{M}=3.0$ (i.e., the upper limit of our {\it Chandra} temperature measurement and the value we found for the sector on the western relic (on$_{\rm{RW}}$), see Tab. \ref{tab:temp_prof}). In this case we obtain $\eta\sim3\times10^{-3}$, still consistent with the DSA framework. Future deeper X-ray observations are therefore required to reduce the uncertainties on the Mach number, and give better constraints on this point. 

Finally, the radio luminosity expected for a $\mathcal{M}=1.7$ shock\footnote{the upper limit of the Mach number we measured in the sector above$_{\rm{RW}}$, where no radio emission has been observed}, using our most optimistic acceleration efficiency ( $\eta=0.05$), is $P_{\rm 1.4~GHz}\sim10^{18}$ W Hz$^{-1}$. This radio power is far below our detection limit. Hence, the lack of radio emission in this sector is still consistent with a DSA scenario.

\section{Summary}\label{sec:conc}
In this paper we presented deep {\it Chandra} (410 ks) and $\it Suzaku$ (180 ks) observations of ZwCl\,0008.8+5215 ($z=0.104$). This galaxy cluster was previously classified as a merging system by means of radio-optical analysis \citep{vanweeren+11,golovich+17} and numerical simulations \citep{kang+12,molnar+broadhurst17}. The previous radio observations revealed the presence of a double radio relic in the east and in the west of the cluster \citep{vanweeren+11}. 

With the new {\it Chandra} observations, we find evidence for the presence of a cold front in the west part of the cluster and, about $2^\prime$ further in the cluster outskirts, a shock. For this shock, we estimate $\mathcal{M}_{S_X}=1.48_{-0.32}^{+0.50}$ and $\mathcal{M}_{T_X}=2.35_{-0.55}^{+0.74}$, from the surface brightness and radial temperature analysis respectively. Additionally, ${\it Suzaku}$ temperature profile suggests a Mach number of $\mathcal{M}_{T_X}=2.02^{+0.74}_{-0.43}$. Given these values, we estimate the shock velocity of $v_{\rm shock,W}=1989^{+509}_{-468}$ km s$^{-1}$, and a consequent time since core passage of $\sim0.3-0.5$ Gyr. The Mach number found with X-ray observations agrees with the one obtained by the radio analysis, assuming diffusive shock acceleration of thermal electrons \citep[i.e. $\mathcal{M}_{\rm RW}=2.4^{+0.4}_{-0.2}$,][]{vanweeren+11}. However, given the large uncertainties on the Mach number, we cannot assert whether this is the leading mechanism for the generation of the relic. 
Also, it remains an open question why the radio relic does not fully trace the full extent of the X-ray shock: we measure ${\rm LLS_{edge,W}\sim1~Mpc}$ and ${\rm LLS_{RW}\sim290~kpc}$ from the X-ray and radio images, respectively. We propose that three radio galaxies, located in the proximity of the relic, might have provided the fossil plasma which has subsequently been re-accelerated. However, no clear connection between the relic and the radio galaxies has been found with the previous radio observation. Further deep and low-frequency observations will be needed to reveal, if present, diffuse and faint radio emission connecting the radio galaxies with the relic \citep[as seen in][for the merging cluster Abell\,34311-3412]{vanweeren+17a}.

In the eastern side of the cluster, where another, longer (i.e. ${\rm LLS_{RE}\sim1.4~Mpc}$), radio relic is observed, we do not find evidence for a shock. We suggest a possible combination of projection effects and position of the relic at the edge of the FOV to explain this. Form the surface brightness profile with $\it Chandra$ we could rule out the presence of shock front with $\mathcal{M}>1.5$, and $\it Suzaku$ temperature measure in the post- and pre-shock regions found $\mathcal{M}_{T_X}=1.54^{+0.65}_{-0.47}$. Both this results disagree with the radio analysis, for which a shock with $\mathcal{M}=2.2^{+0.2}_{-0.1}$ was derived. Further studies, focused on this radio relic, are necessary to better understand its formation scenario.

\begin{acknowledgements}
{\it Acknowledgements:} 
GDG, RJvW and HJAR acknowledge support from the ERC Advanced Investigator programme NewClusters 321271. RJvW acknowledges support of the VIDI research programme with project number 639.042.729, which is financed by the Netherlands Organisation for Scientific Research (NWO).
HA acknowledges the support of NWO via a Veni grant. SRON is supported financially by NWO, the Netherlands Organization for Scientific Research. 
Support for this work was provided by the National Aeronautics and Space Administration through {\it Chandra} Award Numbers GO6-17113X and GO5-14130X issued by the {\it Chandra} X-ray Observatory Center, which is operated by the Smithsonian Astrophysical Observatory for and on behalf of the National Aeronautics Space Administration under contract NAS8-03060. This work was performed under the auspices of the U.S. Department of Energy by Lawrence Livermore National Laboratory under Contract DE-AC52-07NA27344. This research has made use of software provided by the {\it Chandra} X-ray Center (CXC) in the application packages CIAO, ChIPS, and Sherpa. The scientific results reported in this article are based on observations made by the {\it Chandra} X-ray Observatory. 
This research has made use of data obtained from the {\it Suzaku} satellite, a collaborative mission between the space agencies of Japan (JAXA) and the USA (NASA). This research made use of APLpy, an open-source plotting package for Python \citep{2012ascl.soft08017R}.

\end{acknowledgements}

\appendix
\renewcommand\thefigure{\thesection.\arabic{figure}}


\section{MCMC corner plots}\label{apdx:cornerplots}
In this section we present the MCMC ``corner plot'' \citep{foreman-mackey16,foreman-mackey17} for the distribution of the uncertainties in the fitted parameters for the X-ray surface brightness profile across the wedges presented in Figs.~\ref{fig:sb_west}, \ref{fig:sb_east} and \ref{fig:sb_cf}. For all corner plots, contour levels are drawn at $[0.5, 1.0, 1.5, 2.0]\sigma$.
\setcounter{figure}{0} 
\begin{figure*}
\centering
\includegraphics[width=1\textwidth]{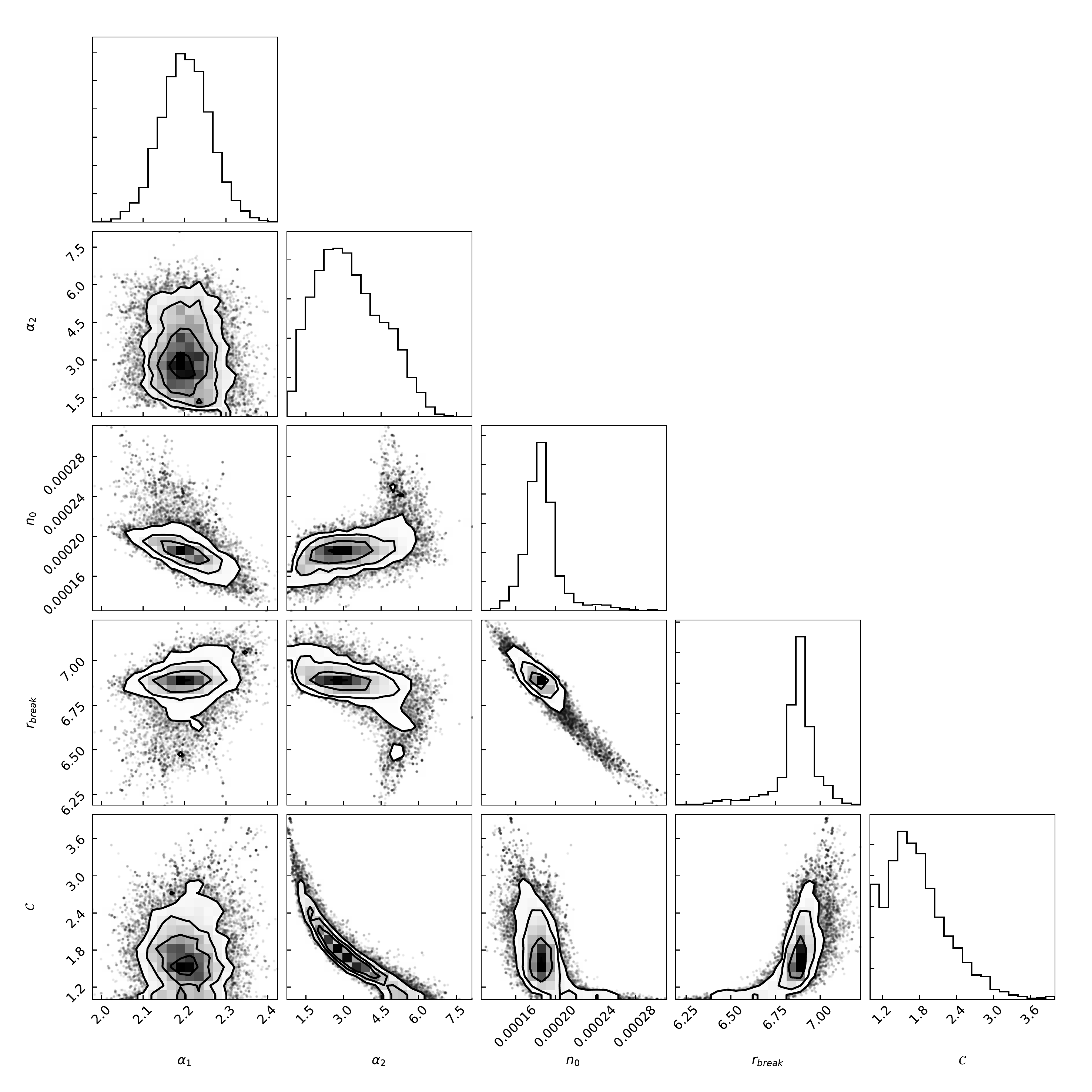}
\caption{The MCMC ``corner plot'' for the X-ray surface brightness profile across the western edge (see top left panel in Fig.~\ref{fig:sb_west})}\label{fig:corner_west}
\end{figure*}

\begin{figure*}
\centering
\includegraphics[width=1\textwidth]{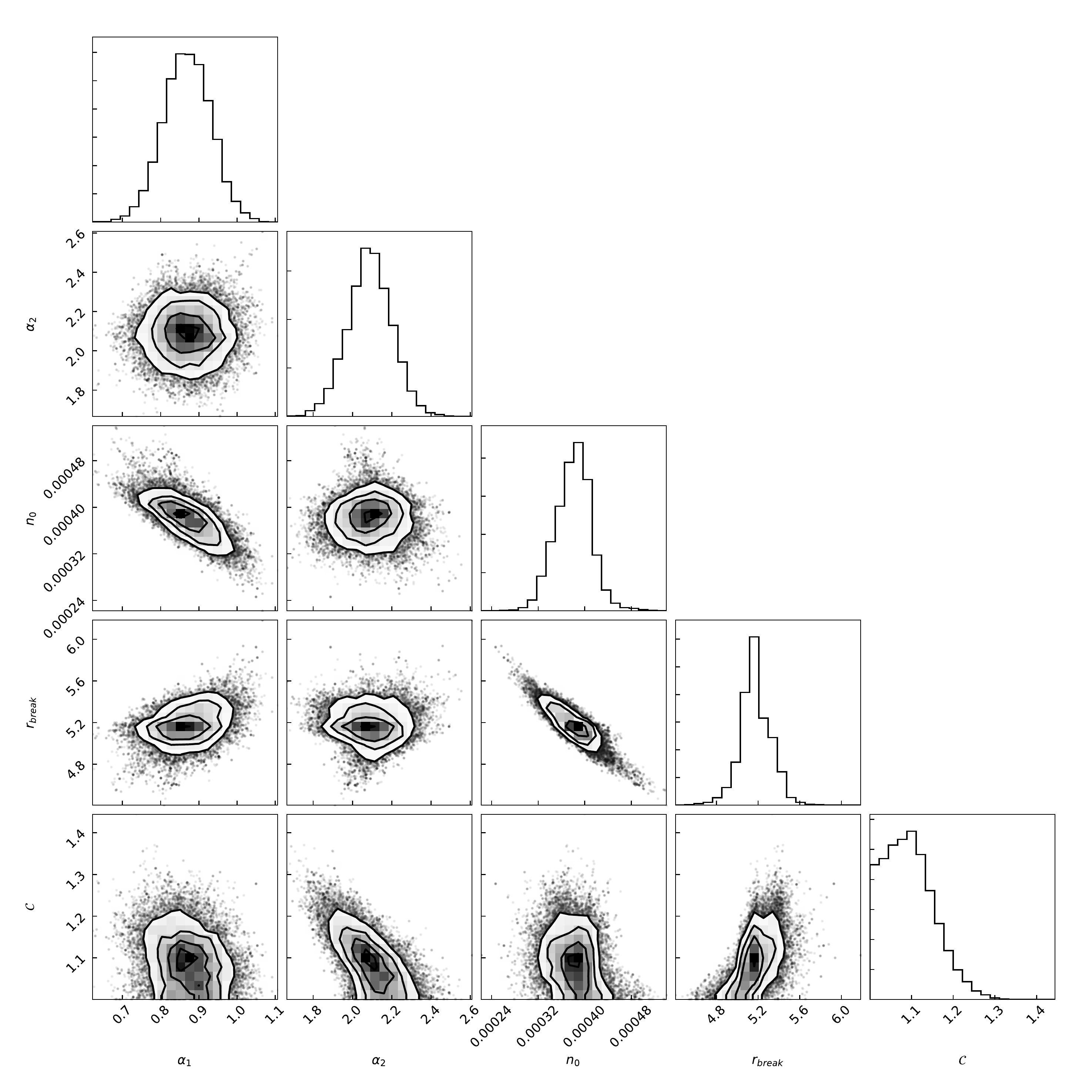}
\caption{The MCMC ``corner plot'' for the X-ray surface brightness profile across the eastern edge (see Fig.~\ref{fig:sb_east}).}\label{fig:corner_east}
\end{figure*}

\begin{figure*}
\centering
\includegraphics[width=1\textwidth]{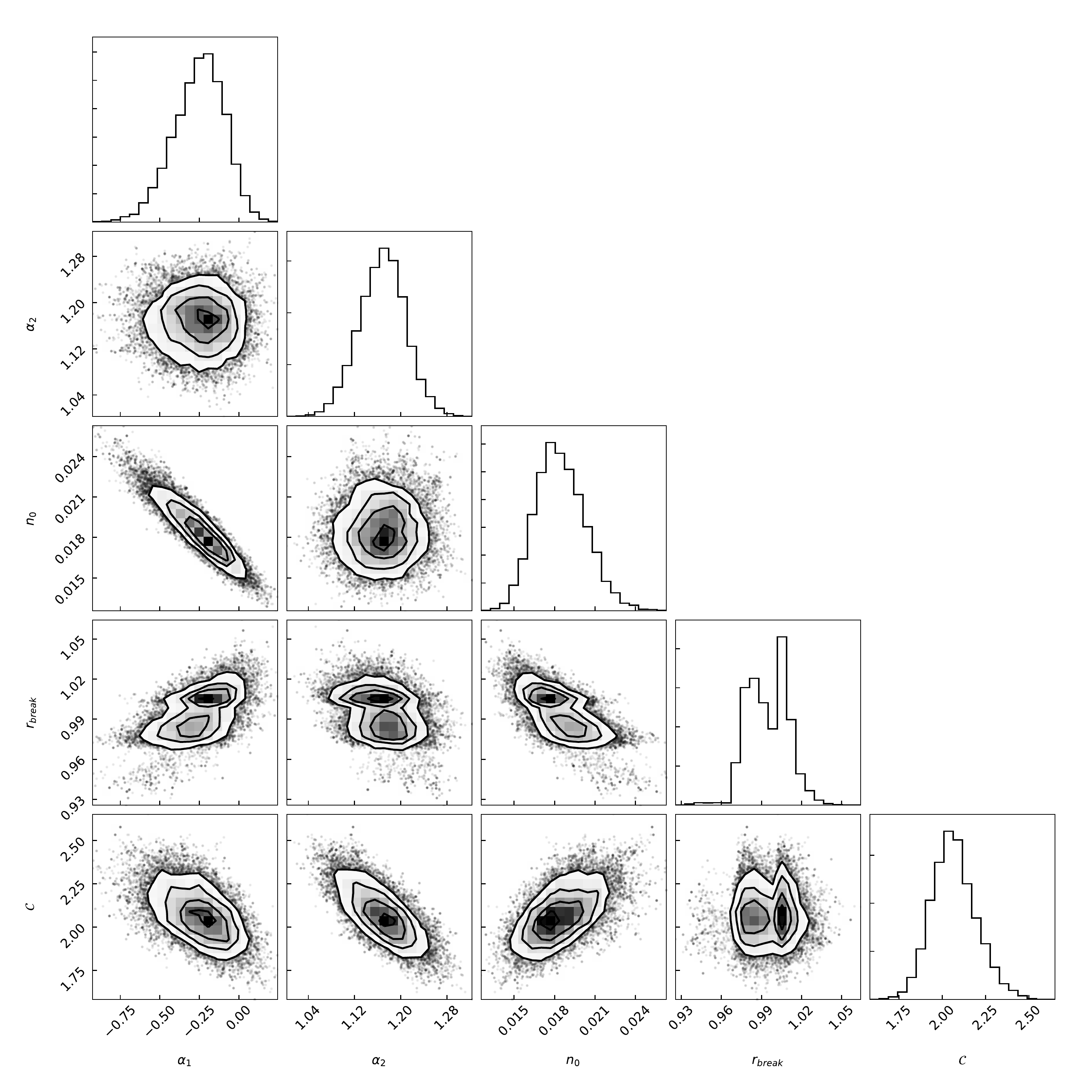}
\caption{The MCMC ``corner plot'' for the X-ray surface brightness profile across the bullet (see Fig.~\ref{fig:sb_cf})}\label{fig:corner_cf}
\end{figure*}

\begin{figure*}
\centering
\includegraphics[width=1\textwidth]{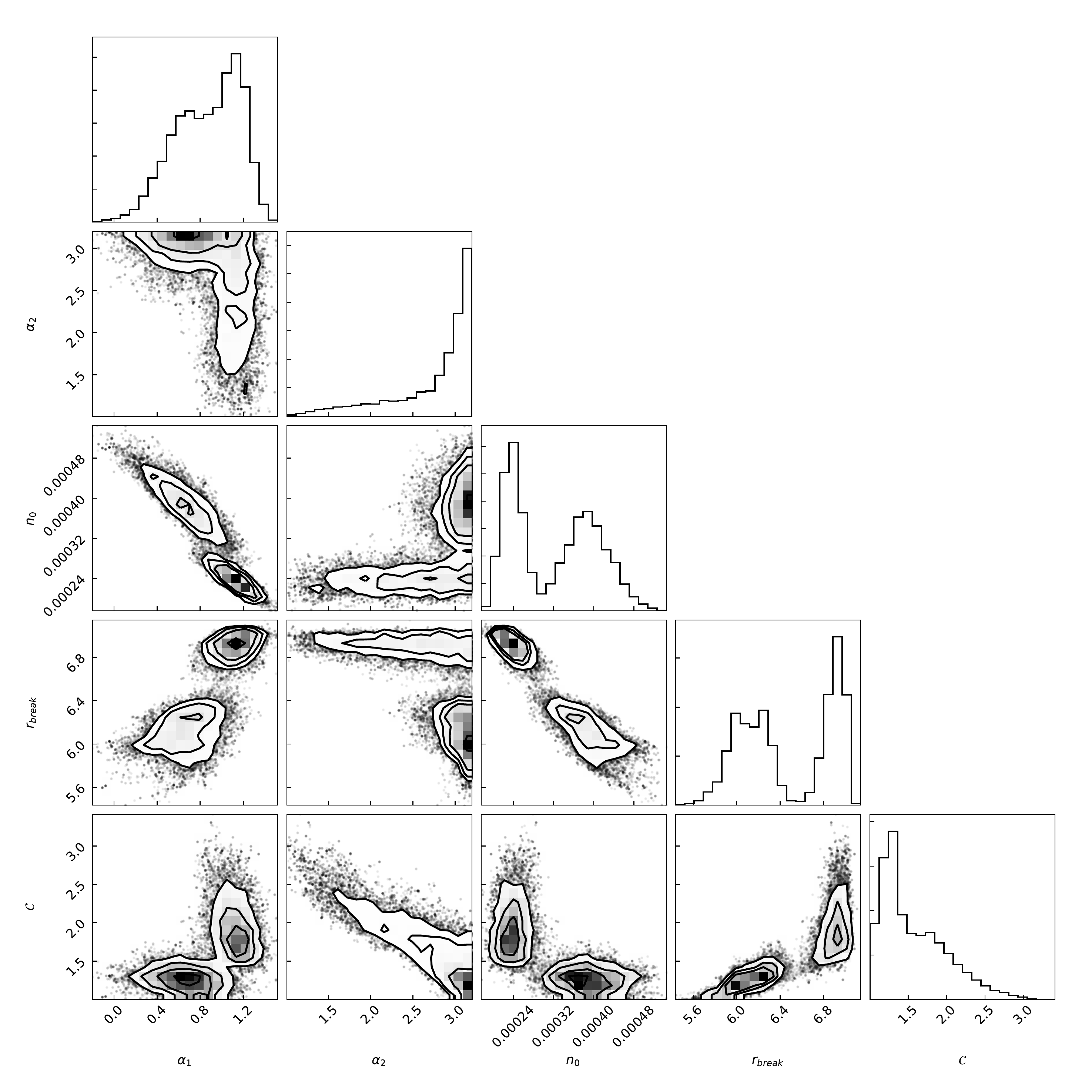}
\caption{The MCMC ``corner plot'' for the X-ray surface brightness profile across the wedge above the western relic (see bottom left panel in Fig.~\ref{fig:sb_west})}\label{fig:corner_abRW}
\end{figure*}

\begin{figure*}
\centering
\includegraphics[width=1\textwidth]{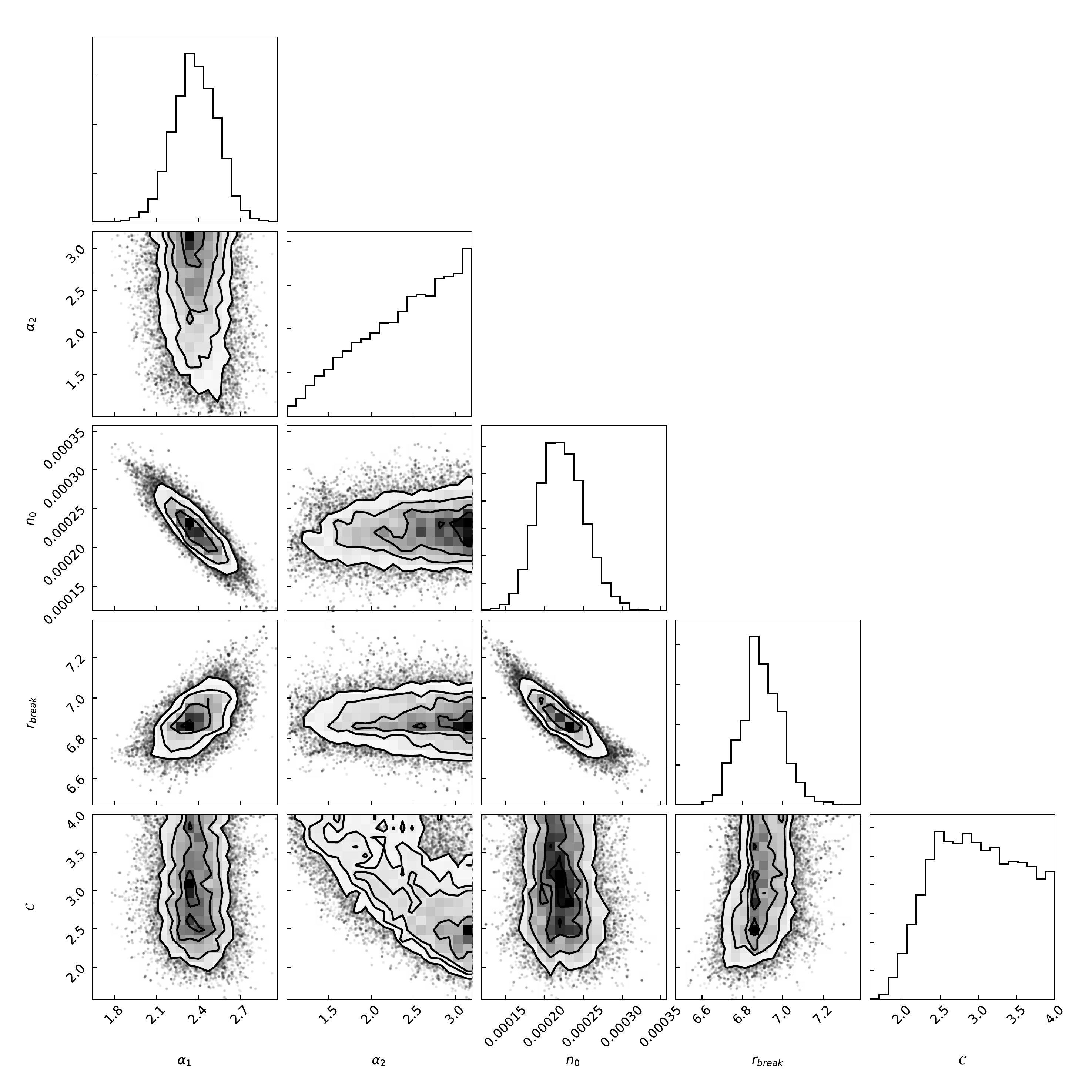}
\caption{The MCMC ``corner plot'' for the X-ray surface brightness profile across the wedge on the western relic (see top right panel in Fig.~\ref{fig:sb_west})}
\label{fig:corner_onRW}
\end{figure*}

\begin{figure*}
\centering
\includegraphics[width=1\textwidth]{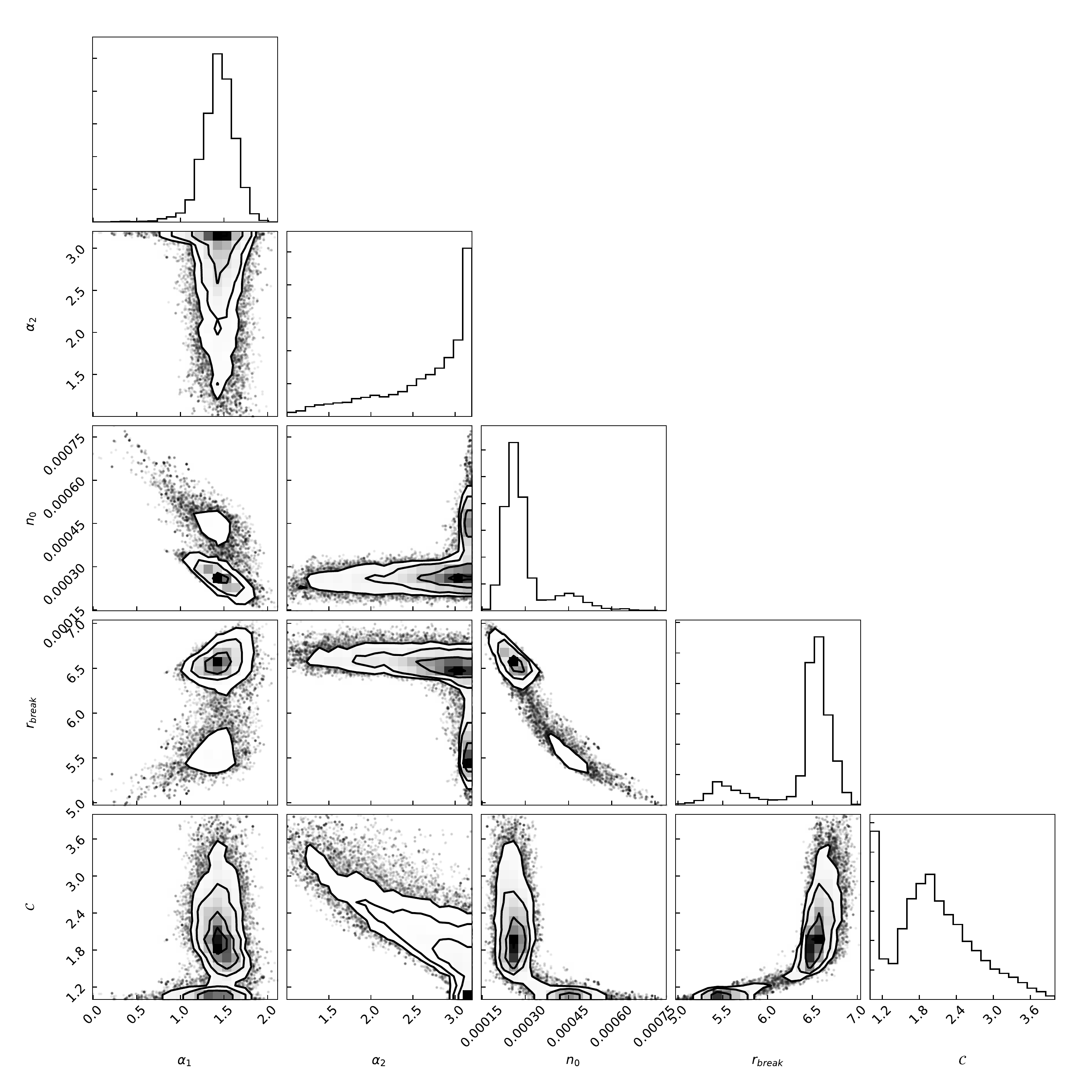}
\caption{The MCMC ``corner plot'' for the X-ray surface brightness profile across the wedge below the western relic (see bottom right panel in Fig.~\ref{fig:sb_west})}\label{fig:corner_blRW}
\end{figure*}


\newpage
\bibliography{biblio.bib}
\end{document}